\UseRawInputEncoding
\documentclass{article}
\usepackage[T1]{fontenc}
\usepackage{textcomp}
\usepackage[utf8]{inputenc}

\usepackage[preprint]{icml2026}

\usepackage{microtype}
\usepackage{hyperref}
\usepackage{graphicx}
\usepackage{wrapfig}

\usepackage[table,dvipsnames]{xcolor}
\usepackage{booktabs,multirow,makecell,tabularx}
\usepackage{siunitx}
\usepackage{amsmath,amssymb,mathtools,amsthm}

\usepackage{caption}
\usepackage{subcaption}
\usepackage{enumitem}
\usepackage[capitalize,noabbrev]{cleveref}

\usepackage{algorithm}
\usepackage{algorithmic}

\usepackage[scaled=0.95]{inconsolata}
\usepackage{listings}
\usepackage[most]{tcolorbox}

\usepackage{xspace}
\usepackage{pifont}   
\usepackage[textsize=tiny]{todonotes}
\usepackage{color}

\theoremstyle{plain}

\theoremstyle{definition}

\theoremstyle{remark}

\newcommand{\cmark}{\textcolor{green!60!black}{\ding{51}}}
\newcommand{\xmark}{\textcolor{red!70!black}{\ding{55}}}

\newcommand{\toolgenesis}{\textsc{Tool~Genesis}\xspace}

\icmltitlerunning{Tool-Genesis: A Task-Driven Tool Creation Benchmark for Self-Evolving Language Agent}

\begin{document}

\twocolumn[
  \icmltitle{Tool-Genesis: A Task-Driven Tool Creation Benchmark for Self-Evolving Language Agent}
    
\icmlsetsymbol{equal}{*}
\icmlsetsymbol{cor}{\textdagger}

\begin{icmlauthorlist}

\icmlauthor{Bowei Xia}{equal,uestc}
\icmlauthor{Mengkang Hu}{equal,hku,xhs,cor}
\icmlauthor{Shijian Wang}{xhs}
\icmlauthor{Jiarui Jin}{xhs}
\icmlauthor{Wenxiang Jiao}{xhs}\\
\icmlauthor{Yuan Lu}{xhs,cor}
\icmlauthor{Kexin Li}{uestc,cor}
\icmlauthor{Ping Luo}{hku}

\end{icmlauthorlist}

\icmlaffiliation{hku}{The University of Hong Kong}
\icmlaffiliation{xhs}{Xiaohongshu Inc.}
\icmlaffiliation{uestc}{UESTC}

\icmlcorrespondingauthor{Mengkang Hu}{mkhu@connect.hku.hk}
\icmlcorrespondingauthor{Kexin Li}{likx@uestc.edu.cn}
\icmlcorrespondingauthor{Yuan Lu}{luyuan3@xiaohongshu.com}

\icmlkeywords{Machine Learning, ICML}

\vskip 0.2in
]



\printAffiliationsAndNotice{}  

\begin{abstract}
Research on self-evolving language agents progresses, increasing attention has focused on their ability to create, adapt, and maintain tools from task requirements.However, existing benchmarks predominantly rely on pre-defined specifications, which limits scalability and hinders true autonomous evolution. While recent studies attempt to dynamically generate tools, they primarily focus on downstream performance, creating a "black box" evaluation that makes it difficult to accurately attribute the causes of failure.To address this, we propose Tool-Genesis, a diagnostic benchmark designed to quantify agent capabilities across multiple dimensions—from interface compliance and functional correctness to downstream utility. It evaluates the ability of agents to construct task-relevant tools solely from abstract requirements (without pre-set specifications) and solve realistic problems.Crucially, we find that even state-of-the-art models struggle to construct precise tool interfaces or executable logic in a one-shot setting. These minor initial flaws are amplified through the pipeline, leading to a precipitous drop in downstream metrics. We hope this benchmark will guide future research toward steering models to synthesize persistent, general-purpose tools capable of addressing broader real-world challenges.\noindent\textbf{Project page:} \url{https://tool-genesis.github.io}
\end{abstract}


\section{Introduction}
\label{sec:introduction}

\begin{table*}[t]
  \centering
  \vspace{-0.4em}
  \small
  \setlength{\tabcolsep}{4.5pt}
  \renewcommand{\arraystretch}{1.15}
  \resizebox{\textwidth}{!}{%
  \begin{tabular}{lccc cccc ccc}
    \toprule
    \multirow{2}{*}{\textbf{Benchmark}} &
    \multicolumn{3}{c}{\textbf{Scale (Reported)}} &
    \multicolumn{4}{c}{\textbf{I. Artifacts Required}} &
    \multicolumn{3}{c}{\textbf{II. Verification Signals}} \\
    \cmidrule(lr){2-4} \cmidrule(lr){5-8} \cmidrule(lr){9-11}
    & \textbf{\makecell{Tool\\Sets}} &
      \textbf{\makecell{Avg\\Tools}} &
      \textbf{\makecell{\#\\Domains}} &
      \textbf{\makecell{No\\Doc}} &
      \textbf{\makecell{Schema\\Gen}} &
      \textbf{\makecell{Reuse \\ Tool}} &
      \textbf{\makecell{Toolset\\($\geq 2$)}} &
      \textbf{\makecell{Held-out\\UnitTests}} &
      \textbf{\makecell{Neg\\Tests}} &
      \textbf{\makecell{GT\\Tools}} \\
    \midrule
CREATOR \cite{qian2023creator} & \makecell{2K} & 1 & \makecell{9}
  & \xmark & \xmark & \cmark & \cmark & \xmark & \xmark & \xmark \\
LATM \cite{cai2024latm} & \makecell{6} & 1 & \makecell{6}
  & \cmark & \xmark & \xmark & \xmark & \xmark & \xmark & \xmark \\
CRAFT \cite{craft} & \makecell{150} & \makecell{3} & 3
  & \xmark & \xmark & \cmark & \cmark & \xmark & \xmark & \xmark \\
TM-Bench \cite{toolmaker} & 15 & 1 & \makecell{4}
  & \cmark & \xmark & \cmark & \xmark & \cmark & \xmark & \xmark \\
SciEvo \cite{2601.07641} & 925 & 6 & 25
      & \cmark & \xmark & \cmark & \xmark & \cmark & \xmark & \xmark \\

    \midrule
    \rowcolor{gray!15}
    \textbf{\textsc{Tool-Genesis} (Ours)} &
      \textbf{86} & \textbf{6} & \textbf{24} &
      \textbf{\cmark} & \textbf{\cmark} & \textbf{\cmark} & \textbf{\cmark} &
      \textbf{\cmark} & \textbf{\cmark} & \textbf{\cmark} \\
    \bottomrule
  \end{tabular}}
  \caption{\textbf{Feature-wise comparison of representative \emph{tool-creation} benchmarks under a strict binary rubric.}
  \textbf{Held-out UnitTests} indicates tests/invocations are \emph{not} available during tool creation and used strictly for evaluation (e.g., TM-Bench).}
  \label{tab:benchmark_comparison}
  \vspace{-0.9em}
\end{table*}

\begin{figure}[t]
  \centering
  \includegraphics[width=\linewidth]{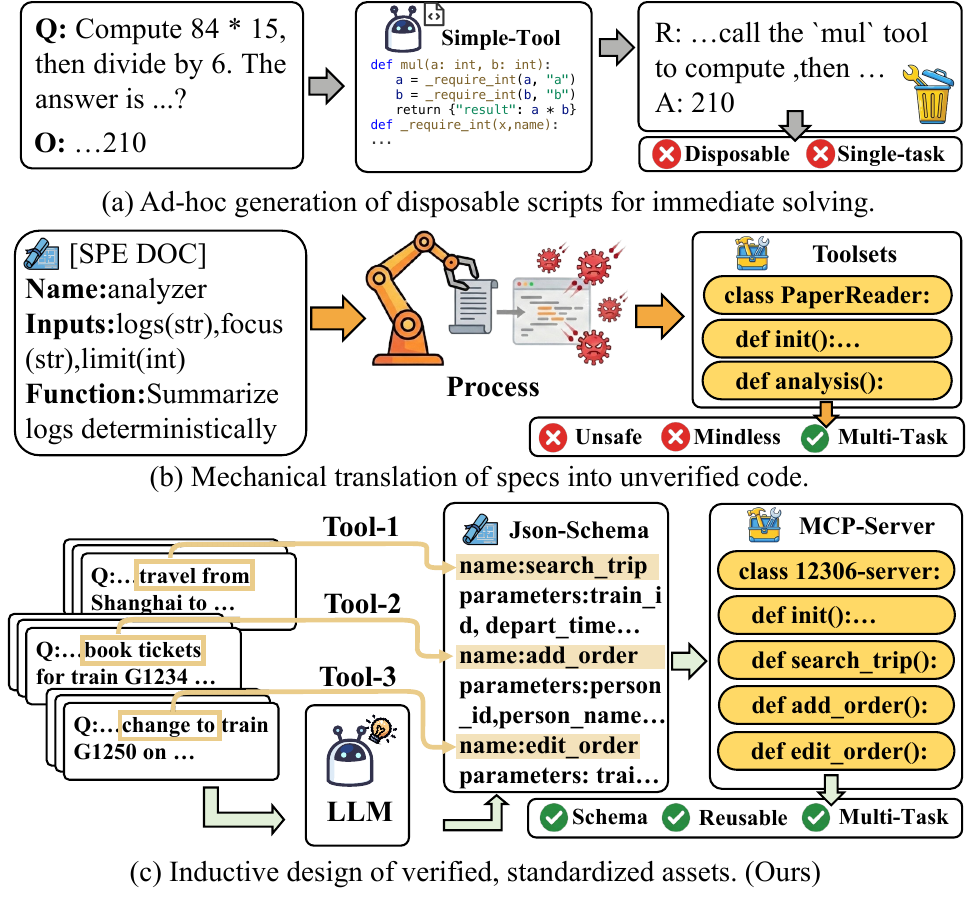}
  \caption{Comparison of tool creation paradigms: (a) Outcome-Driven: Ad-hoc solving with disposable scripts; (b) Code-Centric: Spec-based translation with limited safety; (c) Tool-Genesis(Ours): Inductive design for verified, reusable assets.}
  \label{fig:benchmark_comparison_overview}
  \vspace{-0em} 
\end{figure}

Prior work has established a ``reason--call--execute'' paradigm, typically assuming reliable tool interfaces and schemas, where tools are treated as callable functions with well-defined inputs/outputs and stable semantics.(e.g., \citep{mrkl, yao2022react, toolformer}), with benchmarks further standardizing the evaluation (e.g., \citep{gorilla, toollm, apibank, toolbench2023, stabletoolbench, bfcl, appbench, ultratool, itc, shi2024tool, wang2024executable}).
In this paradigm, tool use is largely reduced to selecting an API, filling arguments, and executing calls under a fixed contract, while success is measured by answer correctness or call-level validity.
In realistic deployments, however, this assumption often breaks due to missing specifications, evolving APIs, uncovered long-tail needs, or execution failures caused by bugs.
Even small interface ambiguities (e.g., optional fields, implicit constraints, undocumented edge cases) can cascade into repeated execution errors and brittle agent behaviors, especially when the task requires multi-step composition across tools.
As a result, agents must evolve from merely \textit{using} tools to \textit{creating}, \textit{adapting}, and \textit{repairing} tools from abstract requirements, and to distilling reusable pipelines into maintainable tool assets---a core mechanism of self-evolving language agents that improve over long-horizon task distributions \citep{zhang2024oscopilot, tan2024cradle}.

Despite steady progress, when the goal is to evaluate this self-evolving capability along the tool dimension under deployment-like constraints, existing benchmarks exhibit three practical disconnects.
First, most evaluations remain \textbf{spec-first}: they assume that interfaces or schemas are directly available, or implicitly rely on high-quality reference specifications. This emphasizes correctness under predefined contracts, while the end-to-end capability of inferring interface contracts from requirements and producing machine-checkable schemas is not systematically measured.
Second, regarding tool organization, many settings primarily evaluate the scale or diversity of tool collections rather than the construction of a \textbf{scenario-closed toolbox}. They often overlook the agent's ability to distill capabilities into a cohesive, maintainable toolset that covers key sub-processes of a specific real-world scenario (e.g., \citealp{craft, trove, ktce, toollibgen, qian-etal-2024-toolink}).
Third, and most critically, evaluation signals are often \textbf{outcome-centric}, creating a \textbf{``black box''} dilemma.
Benchmarks frequently rely on final answers or coarse call-level checks. Even when unit tests are used, their coverage and attribution granularity vary widely.
This makes it difficult to disentangle whether a failure stems from defective tool construction (e.g., invalid schemas, logic bugs) or suboptimal tool utilization strategies, obscuring the specific stage where the error occurred (e.g., \citealp{toolcoder, toolmaker, stabletoolbench, swebench2024, lu2024toolsandbox}).

To address these deployment-facing gaps, we introduce \textbf{Tool-Genesis}, a diagnostic benchmark designed to \textbf{decouple} tool generation from tool utilization.
Unlike spec-first settings, Tool-Genesis evaluates agents under missing or underspecified interfaces, requiring them to infer contracts from abstract requirements, generate machine-checkable schemas, and produce executable implementations that satisfy criteria for reuse and maintenance.
Crucially, our protocol serves as a diagnostic probe: it reveals that even state-of-the-art models struggle to construct precise tool interfaces or logic in a \textbf{one-shot setting}. These minor initial flaws are \textbf{amplified} through the pipeline, leading to precipitous drops in downstream metrics.
By shifting the target from one-off scripts to reusable tool assets, Tool-Genesis evaluates whether agents can continuously distill capabilities to cover a scenario's task distribution.
Finally, we provide a unified, full-lifecycle evaluation protocol that jointly measures compliance, server executability, schema consistency, and functional validation (via explicit negative/boundary tests).
We also introduce an oracle-normalized upper bound to quantify the utility gap between generated tool assets and reference tools.

\begin{itemize}[leftmargin=*]
  \item \textbf{Benchmark setting.}
  We formalize a requirement-driven tool-creation setting that elevates toolsets as reusable assets. It evaluates creation under missing specifications, focusing on the agent's ability to infer schemas and implement executable logic from abstract requirements.
  \item \textbf{Diagnostic evaluation protocol.}
  We provide a full-lifecycle, execution-grounded protocol designed to \textbf{disentangle} failure causes. By incorporating multi-level signals---including compliance, schema fidelity, and explicit negative/boundary unit tests---we enable precise attribution of errors to either tool quality or usage strategy, addressing the ``black box'' issue.
  \item \textbf{Oracle-normalized utility gap.}
  We introduce an oracle-normalized upper-bound comparison to quantify the utility gap between generated tool assets and reference tools under the same task distribution, providing a clearer measure of practical self-evolution capability.
\end{itemize}

\begin{figure*}[!t]
  \centering
  \includegraphics[width=\textwidth]{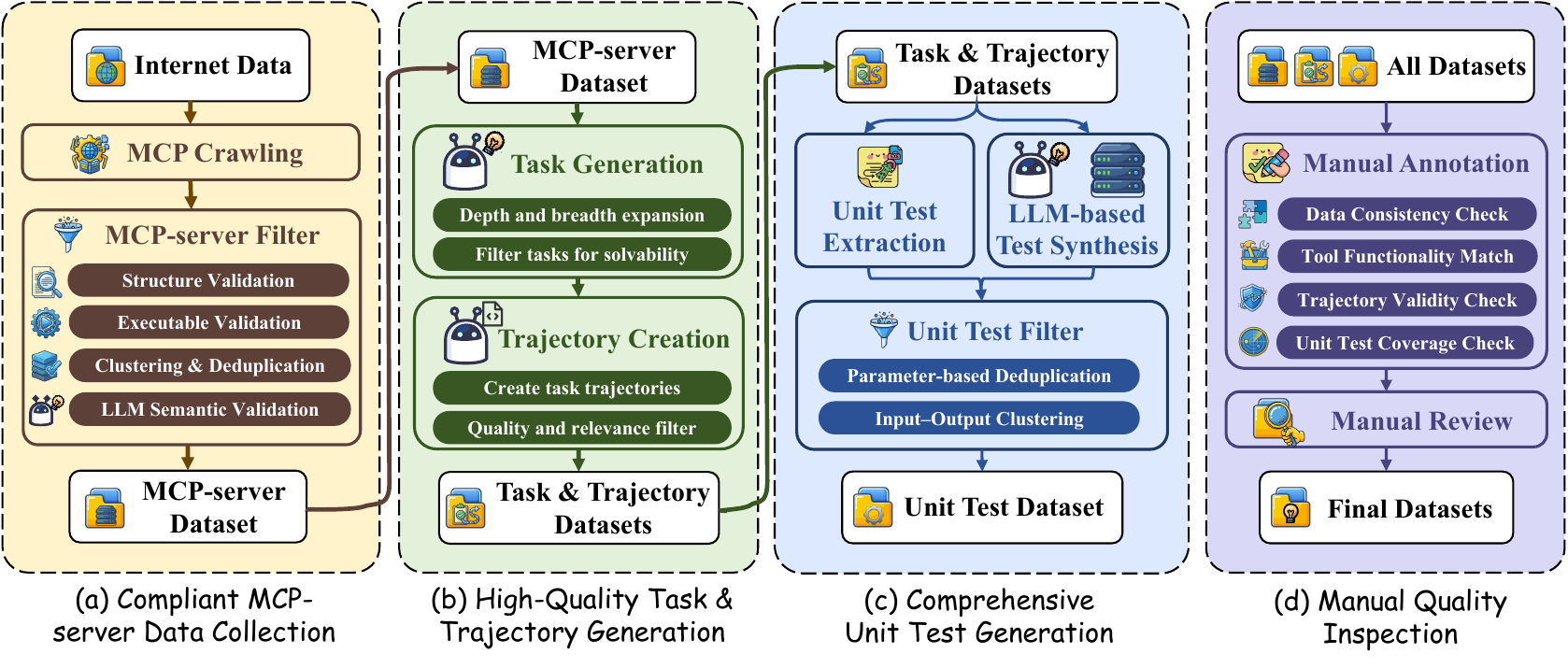}
  \caption{Dataset construction pipeline of \toolgenesis.}
  \label{fig:dataset-construction}
  \vspace{-0.5em}
\end{figure*}

\section{Problem Formalization}
\label{sec:problem_formalization}
\subsection{Task Definition}
\label{sec:task_definition}

We formalize \toolgenesis\ as a conditional generation problem over Model Context Protocol (MCP) interfaces.
Let $\mathcal{X}$ denote the natural-language task description, $\mathcal{S}$ the space of valid MCP interface schemas,
and $\mathcal{E}$ the space of executable server implementations.
An MCP schema $s \in \mathcal{S}$ is represented as an ordered list of atomic tool definitions,
$s = [t_1, \dots, t_K]$, where each tool
$t_k = \langle \eta_k, \phi_k, \delta_k \rangle$ consists of a unique invocation identifier $\eta_k$,
a parameter interface $\phi_k$ specified by a JSON-schema-like typed signature with constraints,
and a natural-language description $\delta_k$ grounding its intended semantics and usage.

We decompose the tool creation process into two coupled prediction phases: \textbf{Tool Interface Prediction} and \textbf{Tool Materialization}.
Formally, the joint probability of producing a schema $s$ and an implementation $e$ given requirement $x$ is factorized as:
\begin{equation}
P_{\theta}(s,e\mid x)
=
\underbrace{P_{\theta}(s\mid x)}_{\text{Interface Prediction}}
\cdot
\underbrace{P_{\theta}(e\mid s)}_{\text{Materialization}}.
\end{equation}
In the first phase, the model predicts an interface schema $\hat{s} = \arg\max_{s \in \mathcal{S}} P_{\theta}(s \mid x)$, specifying the structured tool signatures.
In the second phase, conditioned on a schema $s_{\text{cond}}$, the model materializes an executable server implementation via
$\hat{e} = \arg\max_{e \in \mathcal{E}} P_{\theta}(e \mid s_{\text{cond}})$.
We evaluate this materialization under two settings: \emph{Oracle Materialization}, where $s_{\text{cond}} = s^*$ (ground truth) to isolate engineering capability, and \emph{Cascaded Materialization}, where $s_{\text{cond}} = \hat{s}$ (predicted schema) to assess end-to-end performance.

\subsection{Metrics}
\label{sec:metrics}
We evaluate tool creation with a four-level metric suite (Appendix~\ref{app:metric_details}):  
(i) \textbf{Level 1 (Surface Compliance)} reports Compliance Rate and Server Execution Rate. \textbf{Compliance Rate} measures whether \texttt{list\_tools} returns a parseable, MCP-compliant registry, while \textbf{Server Execution Rate} measures whether the server launches and remains responsive under fixed timeouts.  
(ii) \textbf{Level 2 (Semantic Interface Fidelity)} reports \textbf{Schema-F1}, quantifying schema-level fidelity by aligning predicted and reference tools via bipartite matching and computing an F1 score over tool interfaces.  
(iii) \textbf{Level 3 (Functional Correctness)} reports \textbf{UT$_{\text{soft}}$} and \textbf{UT$_{\text{hard}}$}, which measure the fraction of tools passing predefined \textbf{Unit Tests} under relaxed and strict (boundary/negative) criteria, respectively.  
(iv) \textbf{Level 4 (Downstream Task Utility)} assesses the end-to-end practical efficacy by employing a fixed proxy agent (\texttt{qwen3-14b-instruct}) to solve benchmark tasks equipped with the generated tools. To rigorously isolate tool quality from solver capability, we conduct a parallel control experiment using ground-truth reference tools, with all final outcomes evaluated by an LLM-as-a-Judge. This comparative setup allows us to report an \textbf{Oracle-Normalized Success Rate (SR)}, which quantifies the utility of the synthesized tools relative to the upper-bound performance achieved by the optimal reference implementation under the same experimental conditions.


\newcommand{\TODO}[1]{\textcolor{red}{\textbf{[TODO: #1]}}}

\section{Dataset Construction}
\label{sec:dataset_construction}

This section describes the dataset construction pipeline of \toolgenesis{}. It covers \emph{Compliant MCP-server Data Collection}
(\S\ref{subsec:server_collection}), \emph{High-Quality Task \& Trajectory Generation} (\S\ref{subsec:task_traj_generation}),
\emph{Comprehensive Unit Test Generation} (\S\ref{subsec:unit_test_generation}), and \emph{Manual Quality Inspection}
(\S\ref{subsec:manual_qc}). The overall procedure is illustrated in Figure~\ref{fig:dataset-construction}. Detailed prompts, rubrics, thresholds, and implementation specifics are provided in Appendix~\ref{app:construction_details}.

\subsection{Compliant MCP-server Data Collection}
\label{subsec:server_collection}

\textbf{\textit{MCP Crawling:}}
We collect MCP servers from web sources: (i) MCP aggregators (GLMA, Smithery),
(ii) GitHub search and curated lists, and (iii) HuggingFace (e.g., Toucan) in Aug--Sep 2025.
We keep source links and server metadata (\texttt{server\_name}, description), without mirroring repositories.
We obtain tool registries by launching servers and calling \texttt{list\_tools}, falling back to static
registries/specifications when needed. Schemas are normalized using \texttt{server\_name} and
\texttt{tool\_name} as server/tool IDs.

\vspace{-0.5em}
\textbf{\textit{MCP-server Filtering.}}
We apply a four-stage filtering pipeline to construct a high-quality MCP-server dataset.
(i) \textbf{Structure Validation} enforces that each server exposes a parseable tool registry with
well-formed tool names, descriptions, and input schemas.
(ii) \textbf{Executable Validation} removes servers that cannot be reliably launched or invoked in
a sandboxed environment.
(iii) \textbf{Deduplication \& Clustering} reduces redundancy by grouping servers with similar
schema-level interfaces and retaining a single representative per group.
(iv) \textbf{LLM Semantic Validation} filters servers that require external credentials or exhibit
high sandbox requirements, ensuring safe and self-contained execution.
The remaining servers form the \textbf{MCP-server Dataset} $\mathcal{D}_{srv}$ (\textbf{86} servers).

\subsection{High-Quality Task \& Trajectory Generation}
\label{subsec:task_traj_generation}

\textbf{\textit{Task generation and filtering.}}
We follow a Toucan-style LLM-driven pipeline \citep{toucan} to synthesize tasks.
For each server, an LLM is prompted with tool schemas (and docs, if any) to generate tasks, expanded along
\textbf{breadth} (distinct scenarios and tool subsets) and \textbf{depth} (multi-step tasks with more tool calls).To ensure high data diversity, we further employ a rejection sampling strategy that penalizes redundant tool combinations, forcing the LLM to explore edge cases and rare parameter configurations.
An LLM-as-judge scores candidates on a fixed 1--5 Likert rubric (quality, realism, verifiability, stability) and
assesses \emph{solvability}; we retain only tasks with \textbf{all dimensions $>3$} and \texttt{solvable=true}.

\vspace{-0.5em}

\textbf{\textit{Trajectory generation and filtering.}}
For each retained task, we generate execution trajectories by running an agent in a sandbox(to support servers requiring network access).This sandbox execution ensures that each trajectory is grounded in real-time tool feedback, allowing us to filter out "hallucinated" successful executions that do not reflect actual API behaviors.
We apply lightweight rule-based checks (parseable calls, valid responses) and employ LLM-as-a-judge to verify
consistency and \emph{completion} and penalize redundancy; we retain trajectories with
\textbf{completion and conciseness $>3$} and \texttt{complete=true}, with solvability and completion judged by the LLM.

\subsection{Comprehensive Unit Test Generation}
\label{subsec:unit_test_generation}

\begin{figure}[t]
  \centering
  \includegraphics[width=\linewidth]{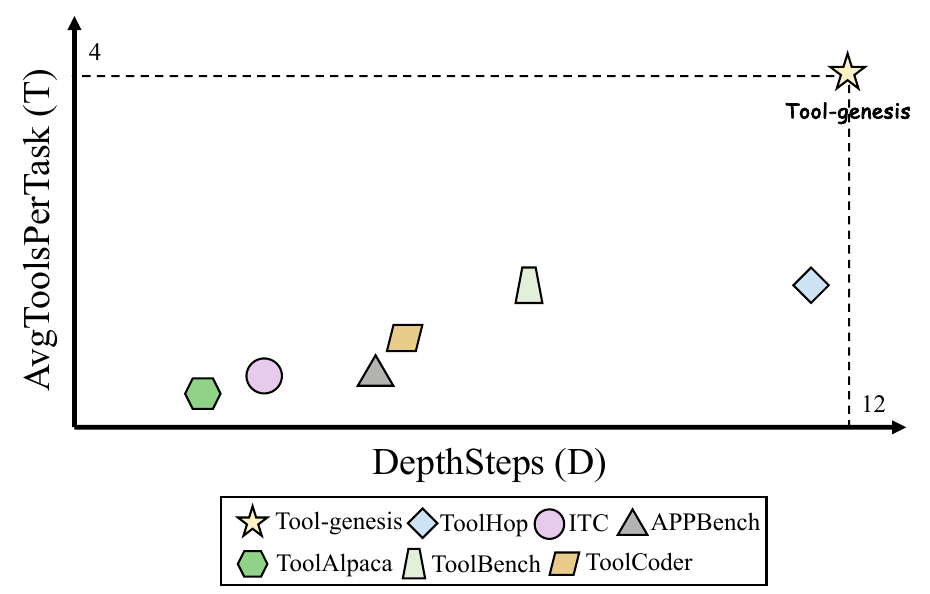}
  \caption{Comparison of benchmarks in terms of task reasoning depth and tool compositionality.}
  \label{fig:task-depth-tool-comp}
\end{figure}

\textbf{\textit{Unit test generation.}}
We extract unit tests from $\mathcal{D}_{traj}$ by converting replayable tool-call steps (executable in the sandbox) into a unified format
(\texttt{tool\_name}, \texttt{inputs}, expected \texttt{outputs}), retaining up to \textbf{100} tests per server for coverage.
When extraction provides insufficient coverage for a tool, we synthesize additional tests with an LLM conditioned on the tool schema,
targeting diverse valid calls.

\vspace{-0.5em}
\textbf{\textit{Unit test filtering.}}
We apply two post-filters for quality and deduplication. (i) \textbf{Parameter-based filtering} removes tests with invalid inputs, type errors,
or disallowed dependencies. (ii) We cluster tests per tool to merge near-duplicates and keep representatives, embedding normalized
\texttt{(input, output)} pairs and merging those with cosine similarity $\ge 0.9$.

\begin{figure*}[t]
  \centering
  \begin{minipage}{0.64\textwidth} 
    \centering
    \includegraphics[width=\textwidth]{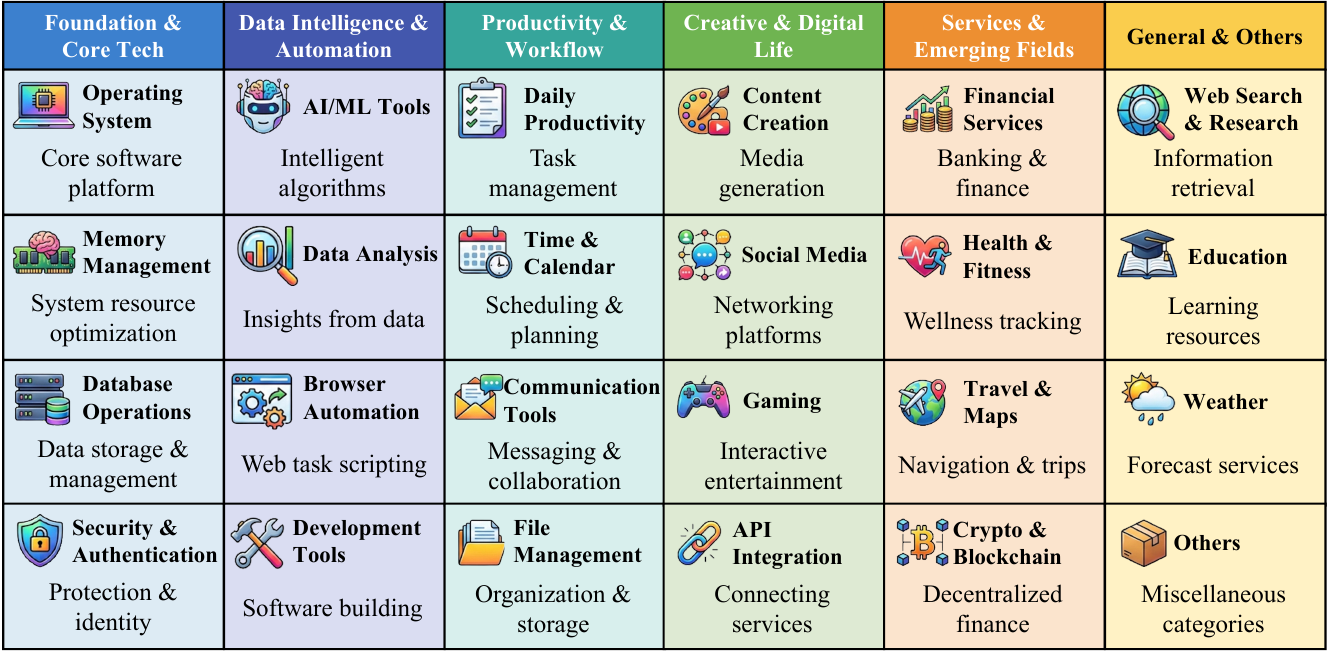}
  \end{minipage}%
  \hspace{0.02\textwidth} 
  \begin{minipage}{0.33\textwidth} 
    \renewcommand{\arraystretch}{1.3}
    \centering
    \begin{tabular}{lc}
    \Xhline{1pt}
    \textbf{Statistic} & \textbf{Number} \\
    \hline
    MCP-servers & 86 \\
    Total Tools & 508 \\
    Domain classes & 24 \\
    Label classes & 18 \\
    \hline
    Unit test & 9441 \\
    Total tasks & 2150 \\
    \hline
    Average task length & 53 \\
    Average step length & 6 \\
    Average tool-using length & 3 \\
    \Xhline{1pt}
    \end{tabular}
  \end{minipage}
  \caption{Overview statistics of \toolgenesis{}.
Left: functional domain coverage of MCP servers across \textbf{24} domain classes.
Right: dataset scale and task/trajectory structure statistics.}

  \label{fig:data_statistics}
  \vspace{-0.6em}
\end{figure*}

\subsection{Manual Quality Inspection}
\label{subsec:manual_qc}

\textbf{\textit{Manual Annotation:}}
(i) \textbf{MCP-server Data Consistency Check}---verify schema/version consistency, stable unique IDs, split integrity, and formatting rules;
(ii) \textbf{Task Tool Functionality Match}---confirm that referenced tools exist in the registry, required arguments are schema-compatible, and the task intent aligns
with documented tool functionality;
(iii) \textbf{Trajectory Validity Check}---ensure trajectories satisfy task constraints with coherent ordering, contain no malformed tool calls, and have no forbidden
dependencies;
(iv) \textbf{Unit Test Coverage Check}---check that tests cover diverse tools and parameter regimes, meet per-server coverage targets, and avoid redundancy.

\vspace{-0.5em}
\textbf{\textit{Manual Review.}}
All instances are manually re-checked by \emph{graduate-level} reviewers (3 annotators over two weeks) in multiple passes.
Reviewers conduct both \emph{instance-level} inspection (task statement, referenced tools, trajectories, and unit tests) and \emph{cross-file} consistency checks (registry $\leftrightarrow$ task $\leftrightarrow$ trajectory $\leftrightarrow$ tests), correcting minor issues when possible and removing samples that violate any requirement.
In particular, they (a) verify that each task is solvable using only the declared toolset and does not rely on hidden assumptions; (b) validate that every tool call conforms to the declared schema (argument names/types, required fields, and return usage); (c) check that trajectories are coherent and free of malformed calls, missing steps, or forbidden external dependencies; and (d) inspect unit tests to ensure they include both positive and negative/boundary cases, span representative parameter regimes, and do not leak answers or duplicate existing tests.
We retain an instance only when at least two annotators independently agree on accept/reject, and the inter-annotator agreement on accept/reject decisions reaches a Cohen's $\kappa$ of $0.85$ \cite{landis1977measurement}.
Finally, after edits and removals, we perform a final end-to-end recheck on the finalized dataset to ensure each retained instance meets all constraints and yields a coherent tool-use process.

\section{Data Analysis}
\label{sec:data_analysis}

\textbf{\textit{Overall scale.}}
Figure~\ref{fig:data_statistics} summarizes the scale of \toolgenesis{}.
After filtering and manual inspection, the dataset contains \textbf{86} executable MCP servers
with \textbf{508} tools, spanning \textbf{24} domain classes.
We collect \textbf{2,150} tasks and \textbf{9,441} unit tests, covering \textbf{18} task label classes.
These statistics provide a concise overview of the benchmark size across servers, tools, tasks, and tests.

\textbf{\textit{Domain coverage.}}
The retained MCP servers span a diverse set of functional domains.
The \textbf{24} domain classes are grouped into six high-level categories, covering
foundational system tools, data intelligence and automation, productivity and workflow utilities,
creative and digital-life applications, service-oriented domains (e.g., finance, health, travel),
and general-purpose tools.
The distribution includes both commonly used domains and a long tail of more specialized categories.

\textbf{\textit{Execution structure.}}
We further examine the execution structure of task trajectories.
Tasks have an average length of \textbf{53} tokens, while trajectories involve
\textbf{6} execution steps on average and invoke \textbf{3} distinct tools per task.
This indicates that many instances require sequential tool invocation rather than single-step execution.
As shown in Figure~\ref{fig:task-depth-tool-comp}, the dataset covers task structures
ranging from simple single-tool interactions to multi-step, multi-tool compositions.

\section{Experiments}\label{sec:experiments}

\newcommand{\mf}[2]{\mbox{#1~\citep{#2}}} 
\newcommand{\nan}{\texttt{NaN}}
\newcommand{\fullsep}{\specialrule{\lightrulewidth}{0pt}{0pt}}

\begin{table*}[!t]
  \centering
  \small
  \setlength{\tabcolsep}{5.5pt}
  \renewcommand{\arraystretch}{1.15}

  \resizebox{\textwidth}{!}{%
  \begin{tabular}{@{}llcccccc@{}}
    \toprule
    \multirow{2}{*}{Model Family} & \multirow{2}{*}{Version}
      & \multicolumn{2}{c}{\textit{L1}}
      & \multicolumn{1}{c}{\textit{L2}}
      & \multicolumn{2}{c}{\textit{L3}}
      & \multicolumn{1}{c}{\textit{L4}} \\
    \cmidrule(lr){3-4}\cmidrule(lr){5-5}\cmidrule(lr){6-7}\cmidrule(lr){8-8}
    & & Compliance $\uparrow$ & Exec.$\uparrow$
      & Schema-F1$\uparrow$
      & UT$_{\text{soft}}\uparrow$ & UT$_{\text{hard}}\uparrow$
      & SR$\uparrow$ \\
    \midrule

    \rowcolor{gray!15}
    \multicolumn{8}{c}{\textbf{Direct}} \\
    \addlinespace[0.25em]

    \multirow{4}{*}{\mf{OpenAI}{openai}}
      & gpt-4o       & 0.779 & 0.209 & 0.175 & 0.089 & 0.049 & 0.153 \\
      & gpt-4.1-mini & 0.686 & 0.318 & 0.293 & 0.127 & 0.088 & 0.199 \\
      & gpt-4.1      & 0.860 & 0.738 & 0.675 & 0.261 & 0.129 & 0.330 \\
      & gpt-5.1      & 0.826 & 0.759 & 0.688 & 0.281 & 0.161 & 0.372 \\
    \fullsep

    \multirow{1}{*}{\mf{Anthropic}{anthropic}}
      & claude-haiku-3.5 & 0.744 & 0.012 & 0.012 & 0.007 & 0.000 & 0.012 \\
    \fullsep

    \multirow{1}{*}{\mf{Google}{google_gemini}}
      & gemini-3-flash-preview & 0.872 & 0.140 & 0.116 & 0.084 & 0.037 & 0.103 \\
    \fullsep

    \multirow{6}{*}{\mf{Qwen3}{qwen}}
      & 235b-a22b-instruct-2507 & 0.884 & 0.333 & 0.320 & 0.143 & 0.108 & 0.287 \\
      & 32b                     & 0.791 & 0.282 & 0.258 & 0.176 & 0.078 & 0.178 \\
      & 30b-a3b-instruct-2507    & 0.884 & 0.256 & 0.250 & 0.093 & 0.025 & 0.218 \\
      & 14b                     & 0.791 & 0.186 & 0.175 & 0.154 & 0.075 & 0.128 \\
      & 8b                      & 0.686 & 0.012 & 0.011 & 0.001 & 0.001 & 0.012 \\
      & 4b                      & 0.651 & 0.000 & 0.000 & 0.000 & 0.000 & 0.000 \\
    \fullsep

    \multirow{1}{*}{\mf{DeepSeek}{deepseek}}
      & deepseek-v3.2 & 0.826 & 0.400 & 0.365 & 0.195 & 0.129 & 0.224 \\
    \fullsep

    \multirow{1}{*}{\mf{MoonshotAI}{moonshot}}
      & Kimi-K2-Instruct & 0.860 & 0.372 & 0.342 & 0.144 & 0.087 & 0.215 \\

    \midrule

    \rowcolor{gray!15}
    \multicolumn{8}{c}{\textbf{Code-Agent}} \\
    \addlinespace[0.25em]

    \multirow{4}{*}{\mf{OpenAI}{openai}}
      & gpt-4o       & 0.802 & 0.531 & 0.456 & 0.237 & 0.117 & 0.226 \\
      & gpt-4.1-mini & 0.651 & 0.906 & 0.640 & 0.278 & 0.144 & 0.344 \\
      & gpt-4.1      & 0.884 & 0.756 & 0.691 & 0.288 & 0.145 & 0.433 \\
      & gpt-5.1      & 0.895 & 0.941 & 0.867 & 0.421 & 0.246 & 0.604 \\
    \fullsep

    \multirow{1}{*}{\mf{Anthropic}{anthropic}}
      & claude-haiku-3.5 & 0.733 & 0.964 & 0.821 & 0.342 & 0.180 & 0.472 \\
    \fullsep

    \multirow{1}{*}{\mf{Google}{google_gemini}}
      & gemini-3-flash-preview & 0.849 & 0.977 & 0.912 & 0.448 & 0.255 & 0.581 \\
    \fullsep

    \multirow{6}{*}{\mf{Qwen3}{qwen}}
      & 235b-a22b-instruct-2507 & 0.686 & 0.971 & 0.722 & 0.363 & 0.212 & 0.472 \\
      & 32b                     & 0.860 & 0.892 & 0.801 & 0.317 & 0.179 & 0.495 \\
      & 30b-a3b-instruct-2507    & 0.744 & 0.913 & 0.694 & 0.336 & 0.187 & 0.410 \\
      & 14b                     & 0.686 & 0.807 & 0.707 & 0.316 & 0.178 & 0.453 \\
      & 8b                      & 0.767 & 0.694 & 0.646 & 0.243 & 0.121 & 0.332 \\
      & 4b                      & 0.651 & 0.326 & 0.285 & 0.127 & 0.058 & 0.181 \\
    \fullsep

    \multirow{1}{*}{\mf{DeepSeek}{deepseek}}
      & deepseek-v3.2 & 0.872 & 0.744 & 0.702 & 0.330 & 0.195 & 0.449 \\
    \fullsep
    
    \multirow{1}{*}{\mf{MoonshotAI}{moonshot}}
      & Kimi-K2 & 0.872 & 0.976 & 0.898 & 0.389 & 0.235 & 0.585 \\

    \bottomrule
  \end{tabular}%
  }

  \caption{Main results under two evaluation paradigms (\textbf{Direct} and \textbf{Code-Agent}) updated with the latest metrics. Metrics follow our four-level evaluation: \textit{L1} surface compliance (Compliance, Server Execution), \textit{L2} semantic interface fidelity (Schema-F1), \textit{L3} functional correctness via unit tests (UT$_{\text{soft}}$, UT$_{\text{hard}}$), and \textit{L4} downstream task utility (Task Success Rate, SR).}
  \label{tab:main-results}
\end{table*}

\subsection{Experimental Setup}
\label{subsec:exp-setup}

\noindent \textbf{Models.}
We evaluate a broad suite of frontier and open-source LLMs, covering closed-source families—OpenAI GPT models (gpt-4o, gpt-4.1-mini, gpt-4.1, gpt-5.1; \citep{openai}), Anthropic Claude (claude-sonnet-4; \citep{anthropic}), and Google Gemini (gemini-3-flash; \citep{google_gemini})—as well as open-source families including Qwen3 (4B/8B/14B/30B-A3B/32B/235B-A22B; \citep{qwen}), DeepSeek (deepseek-v3.2; \citep{deepseek}), and MoonshotAI Kimi (Kimi-K2(-Instruct); \citep{moonshot}).

\noindent \textbf{Inference Strategy.}
On top of this model suite, we run a unified evaluation harness with two inference strategies:
(i) \textbf{Direct}, which performs single-pass generation of TI and TM outputs; and
(ii) \textbf{Code-Agent}, which wraps the same LLM in a ReAct-style agent loop~\citep{yao2022react}.
Specifically, \textbf{Code-Agent} follows a ``think $\rightarrow$ act (tool) $\rightarrow$ observe'' procedure for up to 10 steps, and can invoke sandboxed execution tools to run and validate generated artifacts.

\noindent \textbf{Implementation Details.}
All models are queried through their standard chat APIs using a shared prompt template and fixed decoding settings (max tokens $=40{,}960$, temperature $=0$, top\_p $=1$).
For \textbf{Code-Agent}, sandboxed execution is performed under fixed resource limits and timeouts to support automated artifact validation and downstream task execution.


\subsection{Experimental Results}\label{subsec:exp-results}
Several conclusions can be drawn from TableTable~\ref{tab:main-results}:\textbf{(i) Closed-loop repair yields consistent multi-layer gains and can produce step-changes in end-to-end utility.}
Across most model families, \textit{Code-Agent} improves functional correctness and task success beyond surface executability and schema fidelity.
For Gemini-3-Flash, \textsc{Exec.} increases from $0.140$ to $0.977$, Schema-F1 from $0.116$ to $0.912$, and UT$_{\text{hard}}$ from $0.129$ to $0.726$, resulting in a large SR$_{\text{soft}}$ jump ($0.103 \rightarrow 0.581$).
A similar effect is observed for Qwen3-235B (SR$_{\text{soft}}$: $0.193 \rightarrow 0.622$), indicating that execution feedback is effectively translated into verifiable correctness and downstream utility.\textbf{(ii) High compliance and plausible schemas are necessary but insufficient, revealing a utility-conversion bottleneck.}
Strong upstream signals (L1/L2) do not guarantee downstream success (L4).
Under \textit{Direct}, Qwen3-32B achieves high \textsc{Exec.} ($0.938$) and Schema-F1 ($0.880$), yet SR$_{\text{soft}}$ remains $0.535$.
Conversely, when schema fidelity degrades (e.g., gpt-4.1 under \textit{Direct}), downstream utility collapses accordingly.
These gaps suggest that failures increasingly arise from implementation robustness, boundary handling, and state discipline that are only exposed through execution.\textbf{(iii) Repair benefits are scale-dependent and can reorder models within a family.}
Under \textit{Code-Agent}, larger models achieve higher downstream utility (e.g., Qwen3 SR$_{\text{soft}}$: $0.336$ for 4B vs.\ $0.622$ for 235B).
Notably, model rankings can flip across strategies: Qwen3-32B outperforms 235B under \textit{Direct}, while 235B overtakes under \textit{Code-Agent}.
This indicates that closed-loop tool creation relies on an additional capability---the ability to exploit execution feedback for targeted repair---that is not captured by single-pass generation alone.

\section{Analysis}\label{sec:analysis}


  

\begin{figure*}[t]
  \centering
  \includegraphics[width=\linewidth]{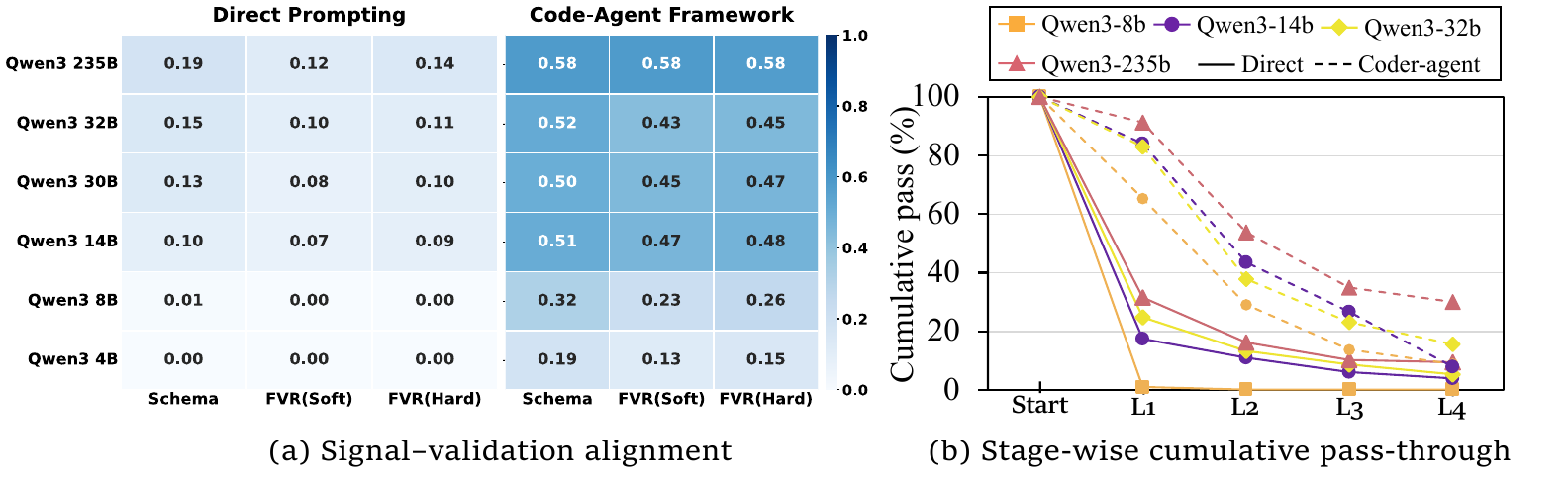}
  \caption{Direct prompting vs.\ code-agent repair on Qwen3 scales: (a) Signal–Validation Alignment (SVA) of Schema-F1 and UT (soft/hard); (b) stage-wise cumulative pass-through across verification stages (Start, L1–L4) for selected scales, with solid/dashed lines denoting the two paradigms.}
  \label{fig:signal_usability}
  \vspace{-0.8em}
\end{figure*}


Beyond aggregate success rates, we analyze tool creation along three axes that align with our lifecycle evaluation.
(i) We measure whether intermediate verification signals are predictive of downstream trajectory validation, and how closed-loop repair changes this predictiveness across model scales (Sec.~\ref{sec:analysis_sva}).
(ii) We localize where failures concentrate along the verification cascade by reporting stage-wise cumulative pass-through from Start to L1--L4, contrasting Direct prompting with Code-Agent repair and identifying the dominant attrition stages (Sec.~\ref{sec:analysis_funnel}).
(iii) We move from inference-time interventions to capability internalization via finetuning, and examine how training reshapes one-shot synthesis under Direct and patch effectiveness under Code-Agent (Sec.~\ref{subsec:explore_sft}).

\subsection{Signal--validation alignment under closed-loop}
\label{sec:analysis_sva}

Figure~\ref{fig:signal_usability}(a) quantifies how \emph{indicative} verification signals are of downstream trajectory validation via a Signal--Validation Alignment (SVA) score (App.~\ref{app:sva_def}).
Under direct prompting, SVA remains low across scales: even at Qwen3-235B, Schema SVA is $0.19$ and UT$_\text{hard}$ is $0.14$, while Qwen3-8B collapses toward zero (Schema $\approx 0.01$, UT$_\text{hard}\approx 0.00$).
Without execution feedback, interface- and verification-level metrics are weak proxies for downstream utility; with closed-loop repair, signal--validation alignment strengthens substantially.
At Qwen3-235B, SVA reaches $0.58$, and the improvement is stage-wise with a threshold around 8B$\rightarrow$14B: Schema increases from $0.32$ to $0.51$, and UT$_\text{hard}$ from $0.26$ to $0.48$.

\subsection{Pass-through along the verification cascade}
\label{sec:analysis_funnel}

Figure~\ref{fig:signal_usability}(b) reports the \emph{cumulative pass-through} (in \%) across Start and L1--L4.
Under direct prompting, Qwen3-8B drops to $1.11\%$ at L1 and $\approx 0.12\%$ by L2--L4; for Qwen3-14B/32B/235B, pass-through is $17.52/24.84/31.60\%$ at L1 and $3.99/5.44/9.47\%$ at L4.
Closed-loop repair lifts early-stage pass-through (L1: $65.34\%$, $84.06\%$, $83.02\%$, $91.36\%$ for 8B/14B/32B/235B) and improves late-stage retention (L4: $8.85\%$, $7.97\%$, $15.69\%$, $30.01\%$), vs.\ $0.12\%$, $3.99\%$, $5.44\%$, and $9.47\%$ under direct prompting.
Once early stages are stabilized, the dominant attrition concentrates at L3/L4, and scaling mainly improves late-stage retention (e.g., L4: $7.97\% \rightarrow 15.69\% \rightarrow 30.01\%$ from 14B to 32B to 235B under code-agent).
Overall, repair primarily reduces early structural failures, leaving deeper validation as the bottleneck.

\vspace{-0.25em}
\subsection{Finetuning exploration: internalizing capability}
\label{subsec:explore_sft}

Prompting and inference-time repair can improve tool creation by injecting demonstrations or execution feedback, but they do not necessarily strengthen the model's intrinsic ability to synthesize reusable tool assets.
We therefore finetune models on \toolgenesis{} and evaluate whether training improves both intermediate verification and downstream task success.
We defer training configurations and implementation details to Appendix~\ref{app:sft_details}.

\begin{figure}[t]
  \centering
  \includegraphics[width=\columnwidth]{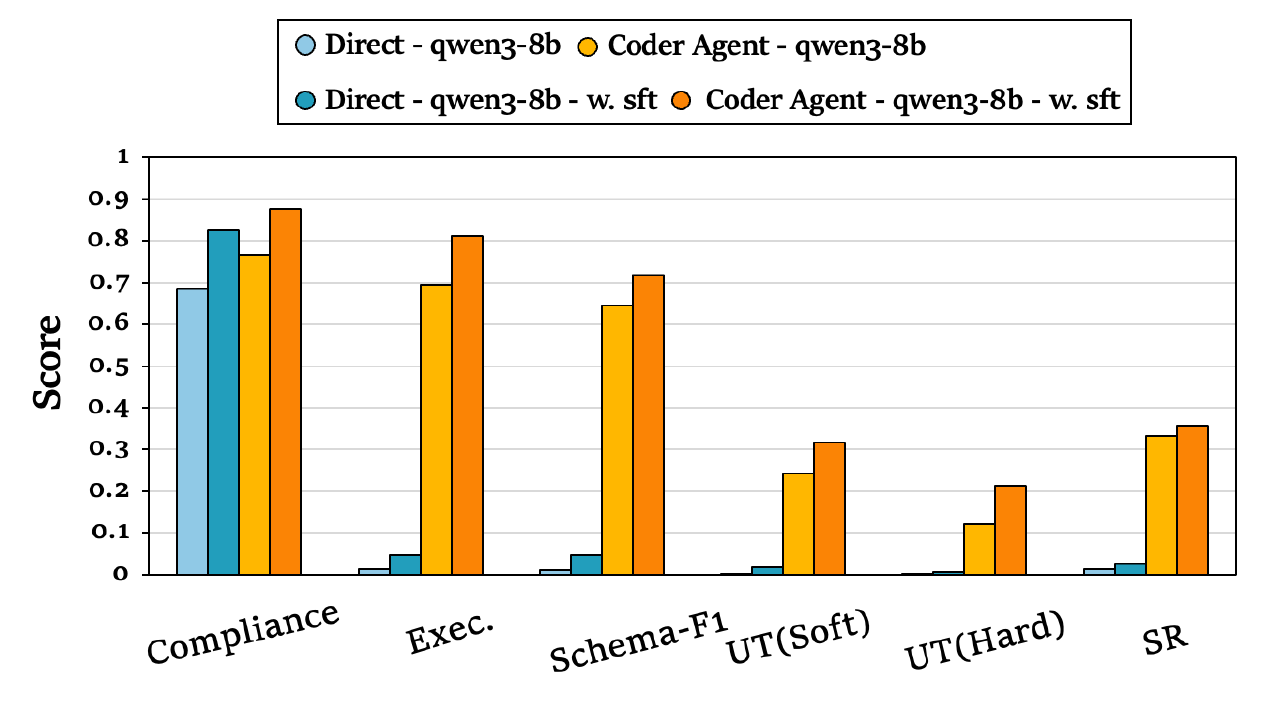}
  \caption{Base vs.\ fine-tuned performance under \textbf{Direct} and \textbf{Code-Agent} evaluation, summarized across our four-layer metric suite (L1--L4).}
  \label{tab:sft-results}
\end{figure}

Finetuning on \toolgenesis{} yields consistent gains across all evaluation layers (Table~\ref{tab:sft-results}), indicating that \toolgenesis{} provides not only a diagnostic benchmark but also an effective training signal for requirement-driven tool creation.
On Qwen3-8B, under the \textbf{Direct} setting, finetuning improves one-shot generation at the interface and schema level: Compliance increases from $0.686$ to $0.826$ ($+0.140$), while Exec.\ rises from $0.012$ to $0.047$ ($+0.035$) and Schema-F1 from $0.011$ to $0.046$ ($+0.035$).
These early-stage gains translate into measurable (though still small) functional validation improvements, with UT$_\text{soft}$ increasing from $0.001$ to $0.017$ ($+0.016$) and UT$_\text{hard}$ from $0.001$ to $0.007$ ($+0.006$), and downstream success SR improving from $0.012$ to $0.026$ ($+0.014$).
Notably, because the Direct baseline is near-zero on executability and testing, absolute gains are more informative than relative ratios in this regime.

Under the \textbf{Code-Agent} setting, finetuning strengthens closed-loop repair and improves how execution feedback is converted into downstream utility.
For Qwen3-8B, Compliance increases from $0.718$ to $0.906$ ($+0.188$), Exec.\ from $0.694$ to $0.777$ ($+0.083$), and Schema-F1 from $0.653$ to $0.736$ ($+0.083$).
Improvements also propagate to functional validation, with UT$_\text{soft}$ increasing from $0.307$ to $0.377$ ($+0.070$) and UT$_\text{hard}$ from $0.456$ to $0.533$ ($+0.077$), yielding a higher SR from $0.336$ to $0.399$ ($+0.063$).
Compared to Direct, gains under Code-Agent are less concentrated at the earliest gates and more evenly reflected in UT and SR, consistent with the interpretation that finetuning improves bug localization and patch quality under a fixed repair budget.

Overall, finetuning complements inference-time repair: it improves one-shot interface/schema synthesis under \textbf{Direct}, and under \textbf{Code-Agent} it increases the effectiveness of execution-triggered debugging, thereby raising the rate at which intermediate verification improvements translate into downstream task success.

\section{Related Work}
\label{sec:related}

Tool-augmented LLMs are commonly evaluated by whether an agent can correctly invoke external tools and APIs to accomplish tasks \citep{mrkl,yao2022react,toolformer,gorilla,apibank,toollm,toolbench,stabletoolbench,bfcl,appbench,ultratool,toolhop,itc,shortcutsbench,taubench,wtueval,agentbench,scienceagentbench,sciagent}. Within this broader line, early tool-creation benchmarks and systems largely treat creation as a task-scoped deliverable and score it with downstream success as a black-box signal \citep{latm,qian-etal-2023-creator,toolmaker}; while task-relevant, such outcome-only evaluation is diagnostically weak because failures conflate requirement misunderstanding, interface/spec errors, implementation bugs, and unsafe or incorrect usage policies. To reduce hallucinated specifications and better reflect real constraints, later work increasingly grounds creation in external references (API docs/specs/repos/knowledge bases) and adds contract-faithfulness signals (e.g., schema/parameter consistency) as intermediate checks, often still reported alongside end-task outcomes \citep{reftool,ktce,qian-etal-2024-toolink,apigen}. As evaluation moves closer to “real execution,” a complementary trend adopts executable and test-driven verification to obtain reproducible correctness signals and clearer attribution \citep{toolcoder,toolmaker,swebench,humaneval,mbpp,apps,evalplus,bigcodebench}; here, a stable execution environment and robust negative/boundary testing become essential to avoid flaky outcomes, prevent accidental overfitting to shallow success cases, and enable regression checking under tool evolution. More recently, tool creation is increasingly framed as building reusable toolsets/toolboxes that support retrieval, composition, maintenance, and updates across a task distribution—often serving knowledge access rather than a single one-off objective \citep{craft,trove,ktce,toollibgen,qian-etal-2024-toolink}. Despite these advances, existing benchmarks remain fragmented: some emphasize task outcomes, others emphasize interface faithfulness or passing tests, and toolset-level reuse/maintenance is rarely evaluated together with executable correctness (including invalid/boundary cases) and downstream utility in a unified, reproducible protocol; in particular, oracle-normalized comparisons that quantify the utility gap between generated tools and ground-truth tools are still under-explored, motivating our benchmark and evaluation design.

\section{Conclusion}
\label{sec:conclusion}

We present \textbf{Tool-Genesis}, a diagnostic benchmark for evaluating tool creation as a first-class capability of self-evolving language agents.
Tool-Genesis departs from spec-first settings by requiring agents to infer tool contracts from abstract requirements, generate machine-checkable schemas, and implement executable logic that can be reused as a scenario-closed toolset.
To avoid outcome-only ``black box'' evaluation, we introduce a full-lifecycle protocol that jointly measures interface compliance, executability, schema fidelity, and functional correctness with explicit negative/boundary unit tests, and we further report an oracle-normalized upper bound to quantify the utility gap between generated and reference tools under the same task distribution.Our empirical findings highlight a key bottleneck: even strong models often fail to produce precise interfaces or correct implementations in a one-shot setting, and these small early-stage defects are amplified through the downstream pipeline.
We hope Tool-Genesis will help the community move beyond ad-hoc, disposable tool use toward \emph{persistent} and \emph{maintainable} tool assets, and enable more targeted progress on tool induction, repair, and verification in realistic deployments.

\section*{Impact Statement}

Authors are \textbf{required} to include a statement of the potential broader
impact of their work, including its ethical aspects and future societal
consequences. This statement should be in an unnumbered section at the end of
the paper (co-located with Acknowledgements -- the two may appear in either
order, but both must be before References), and does not count toward the paper
page limit. In many cases, where the ethical impacts and expected societal
implications are those that are well established when advancing the field of
Machine Learning, substantial discussion is not required, and a simple
statement such as the following will suffice:

``This paper presents work whose goal is to advance the field of Machine
Learning. There are many potential societal consequences of our work, none
which we feel must be specifically highlighted here.''

The above statement can be used verbatim in such cases, but we encourage
authors to think about whether there is content which does warrant further
discussion, as this statement will be apparent if the paper is later flagged
for ethics review.

\nocite{langley00}

\bibliography{example_paper}

@inproceedings{langley00,
 author    = {P. Langley},
 title     = {Crafting Papers on Machine Learning},
 year      = {2000},
 pages     = {1207--1216},
 editor    = {Pat Langley},
 booktitle     = {Proceedings of the 17th International Conference
              on Machine Learning (ICML 2000)},
 address   = {Stanford, CA},
 publisher = {Morgan Kaufmann}
}

@inproceedings{qian-etal-2023-creator,
  title     = {{CREATOR}: Tool Creation for Disentangling Abstract and Concrete Reasoning of Large Language Models},
  author    = {Qian, Cheng and Han, Chi and Fung, Yi and Qin, Yujia and Liu, Zhiyuan and Ji, Heng},
  booktitle = {Findings of the Association for Computational Linguistics: EMNLP 2023},
  year      = {2023},
  month     = dec,
  address   = {Singapore},
  publisher = {Association for Computational Linguistics},
  pages     = {6922--6939},
  doi       = {10.18653/v1/2023.findings-emnlp.462},
  url       = {https://aclanthology.org/2023.findings-emnlp.462/}
}

@inproceedings{qian-etal-2024-toolink,
  title     = {Toolink: Linking Toolkit Creation and Using through Chain-of-Solving on Open-Source Model},
  author    = {Qian, Cheng and Xiong, Chenyan and Liu, Zhenghao and Liu, Zhiyuan},
  booktitle = {Proceedings of the 2024 Conference of the North American Chapter of the Association for Computational Linguistics: Human Language Technologies (Volume 1: Long Papers)},
  year      = {2024},
  month     = jun,
  address   = {Mexico City, Mexico},
  publisher = {Association for Computational Linguistics},
  pages     = {831--854},
  doi       = {10.18653/v1/2024.naacl-long.48},
  url       = {https://aclanthology.org/2024.naacl-long.48/}
}

@misc{toolbench2023,
  title        = {ToolBench: Benchmarking Large Language Models for Tool Manipulation},
  author       = {Anonymous},
  year         = {2023},
  howpublished = {arXiv preprint},
  note         = {Replace with official BibTeX (arXiv / official page).}
}

@misc{swebench2024,
  title        = {{SWE}-bench: Can Language Models Resolve Real-World GitHub Issues?},
  author       = {Anonymous},
  year         = {2024},
  howpublished = {arXiv preprint},
  note         = {Replace with official BibTeX (arXiv / official page).}
}

@misc{mrkl,
  title        = {{MRKL} Systems: A Modular, Neuro-Symbolic Architecture that Combines Large Language Models, External Knowledge Sources and Discrete Reasoning},
  author       = {Karpas, Eli and Shuster, Kurt and Geva, Mor and Schick, Timo and Gupta, Rahul and Eisenstein, Jacob and Berant, Jonathan},
  year         = {2022},
  eprint       = {2205.00445},
  archivePrefix= {arXiv},
  primaryClass = {cs.AI},
  url          = {https://arxiv.org/abs/2205.00445}
}

@misc{toolformer,
  title        = {Toolformer: Language Models Can Teach Themselves to Use Tools},
  author       = {Schick, Timo and Dwivedi-Yu, Jane and Dess{\`i}, Roberto and Raileanu, Roberta and Lombrozo, Tegan and Zettlemoyer, Luke and Cancedda, Nicola and Scialom, Thomas},
  year         = {2023},
  eprint       = {2302.04761},
  archivePrefix= {arXiv},
  primaryClass = {cs.CL},
  url          = {https://arxiv.org/abs/2302.04761}
}

@misc{gorilla,
  title        = {Gorilla: Large Language Model Connected with Massive {APIs}},
  author       = {Patil, Shishir and Zhang, Tianjun and Wang, Xin and Gonzalez, Joseph E.},
  year         = {2023},
  eprint       = {2305.15334},
  archivePrefix= {arXiv},
  primaryClass = {cs.CL},
  url          = {https://arxiv.org/abs/2305.15334}
}

@misc{apibank,
  title        = {{API-Bank}: A Benchmark for Tool-Augmented {LLMs}},
  author       = {Li, Zongxi and Huang, Qingyang and Xu, Hao and Zhang, Zhaozhuo and Wang, Jindong and Li, Ji},
  year         = {2023},
  eprint       = {2304.08244},
  archivePrefix= {arXiv},
  primaryClass = {cs.CL},
  url          = {https://arxiv.org/abs/2304.08244}
}

@misc{toollm,
  title        = {{ToolLLM}: Facilitating Large Language Models to Master 16000+ Real-world {APIs}},
  author       = {Qin, Yujia and Liang, Shihao and Ye, Yining and Zhu, Kunlun and Yan, Lan and Lu, Yaxi and Lin, Yankai and Cong, Xin and Tang, Xiangru and Qian, Bill and Zhao, Sihan and Hong, Lauren and Tian, Runchu and Xie, Ruobing and Zhou, Jie and Gerstein, Mark and Li, Dahai and Liu, Zhiyuan and Sun, Maosong},
  year         = {2023},
  eprint       = {2307.16789},
  archivePrefix= {arXiv},
  primaryClass = {cs.AI},
  url          = {https://arxiv.org/abs/2307.16789}
}

@misc{stabletoolbench,
  title        = {{StableToolBench}: Towards Stable Large-Scale Benchmarking on Tool Learning of Large Language Models},
  author       = {Guo, Zhicheng and Cheng, Sijie and Wang, Hao and Liang, Shihao and Qin, Yujia and Li, Peng and Liu, Zhiyuan and Sun, Maosong and Liu, Yang},
  year         = {2024},
  eprint       = {2403.07714},
  archivePrefix= {arXiv},
  primaryClass = {cs.CL},
  url          = {https://arxiv.org/abs/2403.07714}
}

@misc{bfcl,
  title        = {Berkeley Function Calling Leaderboard: A Live Benchmark for Function Calling Capabilities of Large Language Models},
  author       = {{Berkeley Function Calling Leaderboard (BFCL) Team}},
  year         = {2024},
  howpublished = {OpenReview},
  url          = {https://openreview.net/forum?id=2GmDdhBdDk},
  note         = {Live leaderboard and benchmark for function/tool calling.}
}

@misc{latm,
  title        = {Large Language Models as Tool Makers},
  author       = {Cai, Tianle and Wang, Zhengyang and Chen, Yi and Ma, Ruochen and Li, Ziniu and Zhu, Liwei and Chen, Wanyu and Jiang, Yong and Sun, Maosong and Liu, Zhiyuan},
  year         = {2023},
  eprint       = {2305.17126},
  archivePrefix= {arXiv},
  primaryClass = {cs.CL},
  url          = {https://arxiv.org/abs/2305.17126}
}

@misc{craft,
  title        = {{CRAFT}: Customizing {LLMs} by Creating and Retrieving from Specialized Toolsets},
  author       = {Yuan, Shuai and Chen, Yong and Chen, Yizhou and Chen, Jing and Liu, Zhiyuan},
  year         = {2024},
  eprint       = {2401.04052},
  archivePrefix= {arXiv},
  primaryClass = {cs.CL},
  url          = {https://arxiv.org/abs/2401.04052}
}

@misc{trove,
  title        = {TroVE: A Verifiable and Efficient Framework for Toolset Creation},
  author       = {Wang, Yilin and Zhang, Yichi and Liu, Haoyu and Zhang, Yue},
  year         = {2024},
  eprint       = {2402.04711},
  archivePrefix= {arXiv},
  primaryClass = {cs.SE},
  url          = {https://arxiv.org/abs/2402.04711}
}

@misc{ktce,
  title        = {Knowledge-grounded Tool Creation with Evolution},
  author       = {Zhao, Chen and Wang, Zizhou and Li, Jiapeng and Li, Peng and Zhang, Jian and Zhu, Qingxu and Wang, Jindong and Yuan, Yifei and Gao, Qi and Li, Ji},
  year         = {2025},
  eprint       = {2501.06914},
  archivePrefix= {arXiv},
  primaryClass = {cs.CL},
  url          = {https://arxiv.org/abs/2501.06914}
}

@misc{toollibgen,
  title        = {ToolLibGen: Automated Tool Library Generation with Large Language Models},
  author       = {Huang, Wenqi and Zhuang, Siyuan and Wang, Hao and Guo, Zhicheng and Liu, Zhiyuan and Sun, Maosong and Liu, Yang},
  year         = {2025},
  eprint       = {2505.16956},
  archivePrefix= {arXiv},
  primaryClass = {cs.CL},
  url          = {https://arxiv.org/abs/2505.16956}
}

@misc{toolcoder,
  title        = {ToolCoder: A Holistic Benchmark for Tool Creation},
  author       = {Zhang, Zhe and Liu, Han and Chen, Yixin and Li, Zongxi and Wang, Jindong and Li, Ji},
  year         = {2025},
  eprint       = {2502.11410},
  archivePrefix= {arXiv},
  primaryClass = {cs.SE},
  url          = {https://arxiv.org/abs/2502.11410}
}

@misc{toolmaker,
  title        = {LLM Agents Making Agent Tools},
  author       = {W{\"o}lflein, Georg and Schneider, Jonas and Krenn, Mario},
  year         = {2025},
  eprint       = {2502.11705},
  archivePrefix= {arXiv},
  primaryClass = {cs.CL},
  url          = {https://arxiv.org/abs/2502.11705},
  note         = {Introduces TM-BENCH and the ToolMaker framework.}
}

@misc{reftool,
  title        = {RefTool: Enhancing Model Reasoning with Reference-Guided Tool Creation},
  author       = {Liu, Xiao and Yin, Da and Wu, Zirui and Feng, Yansong},
  year         = {2025},
  eprint       = {2505.21413},
  archivePrefix= {arXiv},
  primaryClass = {cs.CL},
  url          = {https://arxiv.org/abs/2505.21413}
}

@misc{ultratool,
  title        = {Planning, Creation, Usage: Benchmarking LLMs for Comprehensive Tool Utilization in Real-World Complex Scenarios},
  author       = {Huang, Shijue and Zhong, Wanjun and Lu, Jianqiao and Zhu, Qi and Gao, Jiahui and Liu, Weiwen and Hou, Yutai and Zeng, Xingshan and Wang, Yasheng and Shang, Lifeng and Jiang, Xin and Xu, Ruifeng and Liu, Qun},
  year         = {2024},
  eprint       = {2401.17167},
  archivePrefix= {arXiv},
  primaryClass = {cs.CL},
  url          = {https://arxiv.org/abs/2401.17167}
}

@misc{toucan,
  title        = {TOUCAN: Synthesizing 1.5M Tool-Agentic Data from Real-World MCP Environments},
  author       = {Xu, Zhangchen and Meza Soria, Adriana and Tan, Shawn and Roy, Anurag and Agrawal, Ashish Sunil and Poovendran, Radha and Panda, Rameswar},
  year         = {2025},
  eprint       = {2510.01179},
  archivePrefix= {arXiv},
  primaryClass = {cs.LG},
  doi          = {10.48550/arXiv.2510.01179},
  url          = {https://arxiv.org/abs/2510.01179}
}

@misc{itc,
  title         = {International Tool Calling (ITC): A Benchmark for Cross-Lingual and Cross-Regional Tool Use},
  author        = {Zhang, Zuoyu and Zhu, Yancheng and Chen, Xiaojun and Liu, Ziqi and Zhang, Qin},
  year          = {2025},
  howpublished  = {OpenReview (ICLR 2026 submission, withdrawn)},
  url           = {https://openreview.net/forum?id=5OCbI4bJQ7}
}

@misc{toolhop,
  title         = {ToolHop: A Query-Driven Benchmark for Multi-hop Tool Use},
  author        = {Anonymous},
  year          = {2025},
  eprint        = {2501.02506},
  archivePrefix = {arXiv},
  primaryClass  = {cs.CL},
  url           = {https://arxiv.org/abs/2501.02506}
}

@misc{apigen,
  title         = {{APIGen}: Automated Pipeline for {API} Function Calling Data Generation},
  author        = {Wang, Yilun and others},
  year          = {2024},
  eprint        = {2409.01406},
  archivePrefix = {arXiv},
  primaryClass  = {cs.CL},
  url           = {https://arxiv.org/abs/2409.01406}
}

@article{landis1977measurement,
  title        = {The Measurement of Observer Agreement for Categorical Data},
  author       = {Landis, J. Richard and Koch, Gary G.},
  journal      = {Biometrics},
  volume       = {33},
  number       = {1},
  pages        = {159--174},
  year         = {1977},
  publisher    = {International Biometric Society}
}

@inproceedings{appbench,
  title     = "{A}pp{B}ench: Planning of Multiple {API}s from Various {APP}s for Complex User Instruction",
  author    = "Wang, Hongru and Wang, Rui and Xue, Boyang and Xia, Heming and Cao, Jingtao and Liu, Zeming and Pan, Jeff Z. and Wong, Kam-Fai",
  editor    = "Al-Onaizan, Yaser and Bansal, Mohit and Chen, Yun-Nung",
  booktitle = "Proceedings of the 2024 Conference on Empirical Methods in Natural Language Processing",
  month     = nov,
  year      = "2024",
  address   = "Miami, Florida, USA",
  publisher = "Association for Computational Linguistics",
  url       = "https://aclanthology.org/2024.emnlp-main.856/",
  doi       = "10.18653/v1/2024.emnlp-main.856",
  pages     = "15322--15336"
}

@misc{openai,
  title        = {Update to {GPT-5} System Card: {GPT-5.2}},
  author       = {{OpenAI}},
  year         = {2025},
  month        = dec,
  url          = {https://cdn.openai.com/pdf/3a4153c8-c748-4b71-8e31-aecbde944f8d/oai_5_2_system-card.pdf},
  note         = {System card (PDF). Accessed: 2026-01-22}
}

@misc{anthropic,
  title        = {Claude Opus 4 \& Claude Sonnet 4: System Card},
  author       = {{Anthropic}},
  year         = {2025},
  month        = jul,
  url          = {https://www.anthropic.com/claude-4-system-card},
  note         = {System card (PDF). Accessed: 2026-01-22}
}

@misc{google_gemini,
  title        = {Gemini 3 Flash: Model Card},
  author       = {{Google DeepMind}},
  year         = {2025},
  month        = dec,
  url          = {https://storage.googleapis.com/deepmind-media/Model-Cards/Gemini-3-Flash-Model-Card.pdf},
  note         = {Model card (PDF). Accessed: 2026-01-22}
}

@misc{qwen,
  title        = {Qwen3 Technical Report},
  author       = {Yang, An and others},
  year         = {2025},
  eprint       = {2505.09388},
  archivePrefix= {arXiv},
  primaryClass = {cs.CL},
  url          = {https://arxiv.org/abs/2505.09388}
}

@misc{deepseek,
  title        = {DeepSeek-V3 Technical Report},
  author       = {{DeepSeek-AI} and others},
  year         = {2024},
  eprint       = {2412.19437},
  archivePrefix= {arXiv},
  primaryClass = {cs.CL},
  url          = {https://arxiv.org/abs/2412.19437}
}

@misc{moonshot,
  title        = {Kimi K2: Open Agentic Intelligence},
  author       = {{Kimi Team}},
  year         = {2025},
  eprint       = {2507.20534},
  archivePrefix= {arXiv},
  primaryClass = {cs.CL},
  url          = {https://arxiv.org/abs/2507.20534}
}

@inproceedings{yao2022react,
  title     = {ReAct: Synergizing Reasoning and Acting in Language Models},
  author    = {Yao, Shunyu and Zhao, Jeffrey and Yu, Dian and Du, Nan and Shafran, Izhak and Narasimhan, Karthik and Cao, Yuan},
  booktitle = {International Conference on Learning Representations (ICLR)},
  year      = {2023},
  url       = {https://arxiv.org/abs/2210.03629}
}

@misc{toolbench,
  title         = {ToolBench: Towards Benchmarking Large Language Models on Tool Use},
  author        = {Qin, Yujia and Liang, Shihao and Ye, Yining and Zhu, Kunlun and Yan, Lan and Lu, Yaxi and Lin, Yankai and Cong, Xin and Tang, Xiangru and others},
  year          = {2023},
  eprint        = {2305.16504},
  archivePrefix = {arXiv},
  primaryClass  = {cs.CL},
  doi           = {10.48550/arXiv.2305.16504},
  url           = {https://arxiv.org/abs/2305.16504}
}

@misc{shortcutsbench,
  title         = {ShortcutsBench: A Large-Scale Real-world Benchmark for API-based Agents},
  author        = {Shen, Haoyang and Li, Yinuo and others},
  year          = {2024},
  eprint        = {2407.00132},
  archivePrefix = {arXiv},
  primaryClass  = {cs.CL},
  doi           = {10.48550/arXiv.2407.00132},
  url           = {https://arxiv.org/abs/2407.00132}
}

@misc{taubench,
  title         = {$\tau$-bench: A Benchmark for Tool-Agent User Interaction in Real-World Scenarios},
  author        = {Liu, Xiao and others},
  year          = {2024},
  eprint        = {2406.12045},
  archivePrefix = {arXiv},
  primaryClass  = {cs.CL},
  doi           = {10.48550/arXiv.2406.12045},
  url           = {https://arxiv.org/abs/2406.12045}
}

@misc{wtueval,
  title         = {WTU-Eval: A Unified Evaluation Framework for Tool Use},
  author        = {Wang, Zhen and others},
  year          = {2024},
  eprint        = {2407.12823},
  archivePrefix = {arXiv},
  primaryClass  = {cs.CL},
  doi           = {10.48550/arXiv.2407.12823},
  url           = {https://arxiv.org/abs/2407.12823}
}

@misc{agentbench,
  title         = {AgentBench: Evaluating LLMs as Agents},
  author        = {Liu, Tianyu and others},
  year          = {2023},
  eprint        = {2308.03688},
  archivePrefix = {arXiv},
  primaryClass  = {cs.AI},
  doi           = {10.48550/arXiv.2308.03688},
  url           = {https://arxiv.org/abs/2308.03688}
}

@misc{scienceagentbench,
  title         = {ScienceAgentBench: Benchmarking Language Agents for Scientific Problem Solving},
  author        = {Zhang, Yichi and others},
  year          = {2024},
  eprint        = {2410.05080},
  archivePrefix = {arXiv},
  primaryClass  = {cs.AI},
  doi           = {10.48550/arXiv.2410.05080},
  url           = {https://arxiv.org/abs/2410.05080}
}

@misc{sciagent,
  title         = {SciAgent: Tool-augmented Language Models for Scientific Reasoning},
  author        = {Wang, Yubo and others},
  year          = {2024},
  eprint        = {2402.11451},
  archivePrefix = {arXiv},
  primaryClass  = {cs.CL},
  doi           = {10.48550/arXiv.2402.11451},
  url           = {https://arxiv.org/abs/2402.11451}
}

@misc{swebench,
  title         = {{SWE}-bench: Can Language Models Resolve Real-World GitHub Issues?},
  author        = {Jimenez, Carlos E. and Yang, John and Wettig, Alexander and Yao, Shunyu and Pei, Kexin and Press, Ofir and Narasimhan, Karthik},
  year          = {2023},
  eprint        = {2310.06770},
  archivePrefix = {arXiv},
  primaryClass  = {cs.SE},
  doi           = {10.48550/arXiv.2310.06770},
  url           = {https://arxiv.org/abs/2310.06770}
}

@misc{humaneval,
  title         = {Evaluating Large Language Models Trained on Code},
  author        = {Chen, Mark and Tworek, Jerry and Jun, Heewoo and Yuan, Qiming and Pinto, Henrique Ponde de Oliveira and Kaplan, Jared and Edwards, Harri and others},
  year          = {2021},
  eprint        = {2107.03374},
  archivePrefix = {arXiv},
  primaryClass  = {cs.LG},
  doi           = {10.48550/arXiv.2107.03374},
  url           = {https://arxiv.org/abs/2107.03374}
}

@misc{mbpp,
  title         = {Program Synthesis with Large Language Models},
  author        = {Austin, Jacob and Odena, Augustus and Nye, Maxwell and Bosma, Maarten and Michalewski, Henryk and Dohan, David and Jiang, Ellen and Cai, Carrie and Terry, Michael and others},
  year          = {2021},
  eprint        = {2108.07732},
  archivePrefix = {arXiv},
  primaryClass  = {cs.PL},
  doi           = {10.48550/arXiv.2108.07732},
  url           = {https://arxiv.org/abs/2108.07732},
  note          = {Introduces MBPP (Mostly Basic Python Problems).}
}

@misc{apps,
  title         = {Measuring Coding Challenge Competence With APPS},
  author        = {Hendrycks, Dan and Basart, Steven and Kadavath, Saurav and Mazeika, Mantas and Arora, Akul and Guo, Ethan and Burns, Collin and Puranik, Samir and He, Horace and Song, Dawn and Steinhardt, Jacob},
  year          = {2021},
  eprint        = {2105.09938},
  archivePrefix = {arXiv},
  primaryClass  = {cs.LG},
  doi           = {10.48550/arXiv.2105.09938},
  url           = {https://arxiv.org/abs/2105.09938}
}

@misc{evalplus,
  title         = {Is Your Code Generated by ChatGPT Really Correct? Rigorous Evaluation of Large Language Models for Code Generation},
  author        = {Liu, Jiawei and Xia, Chun and Zhang, Lingming and others},
  year          = {2023},
  eprint        = {2305.01210},
  archivePrefix = {arXiv},
  primaryClass  = {cs.SE},
  doi           = {10.48550/arXiv.2305.01210},
  url           = {https://arxiv.org/abs/2305.01210}
}

@misc{bigcodebench,
  title         = {BigCodeBench: Benchmarking Code Generation with Diverse Function Calls and Complex Instructions},
  author        = {Zhuo, Terry Yue and others},
  year          = {2024},
  eprint        = {2406.15877},
  archivePrefix = {arXiv},
  primaryClass  = {cs.SE},
  doi           = {10.48550/arXiv.2406.15877},
  url           = {https://arxiv.org/abs/2406.15877}
}

@article{qian2023creator,
  title={CREATOR: A Benchmark for Tool Creation and Execution},
  author={Qian, X. and Zhang, Y. and Li, Z. and Wang, L.},
  journal={Journal of Artificial Intelligence Research},
  volume={69},
  pages={1--23},
  year={2023},
  publisher={AI Research Institute}
}

@article{cai2024latm,
  title={LATM: A Benchmark for Latent Tool Creation},
  author={Cai, Z. and Liu, Y. and Zhang, W. and Wang, Y.},
  journal={Proceedings of the 2024 International Conference on Machine Learning (ICML)},
  pages={234--245},
  year={2024},
  publisher={ICML}
}

@article{2601.07641,
  title={SciEvo: A Benchmark for Scientific Evolution of Tools},
  author={Zhang, Y. and Liu, H. and Huang, T. and Chen, Q.},
  journal={arXiv preprint arXiv:2601.07641},
  year={2025}
}

@inproceedings{wang2024executable,
  title={Executable Code Actions Elicit Better LLM Agents},
  author={Wang, Xingyao and Chen, Yangyi and Yuan, Lifan and Zhang, Yizhe and Li, Yunzhu and Peng, Hao and Ji, Heng},
  booktitle={Forty-first International Conference on Machine Learning (ICML)},
  year={2024}
}

@article{shi2024tool,
  title={Tool Learning in the Wild: Empowering Language Models as Automatic Tool Agents},
  author={Shi, Zhengliang and Gao, Shen and Yan, Lingyong and Feng, Yue and Chen, Xiuyi and Chen, Zhumin and Yin, Dawei and Verberne, Suzan and Ren, Zhaochun},
  journal={arXiv preprint arXiv:2405.16533},
  year={2024}
}

@article{zhang2024oscopilot,
  title={OS-Copilot: Towards Generalist Computer Agents with Self-Improvement},
  author={Wu, Zhiyong and Han, Chengcheng and Ding, Zichen and Weng, Zhenmin and Liu, Zhoumianze and Yao, Shunyu and Yu, Tao and Kong, Lingpeng},
  journal={arXiv preprint arXiv:2402.07456},
  year={2024}
}

@article{tan2024cradle,
  title={Cradle: Empowering Foundation Agents Towards General Computer Control},
  author={Tan, Weihao and Ding, Ziluo and Zhang, Wentao and Li, Boyu and Zhou, Bohan and Yue, Junpeng and others},
  journal={arXiv preprint arXiv:2403.03186},
  year={2024}
}

@article{lu2024toolsandbox,
  title={ToolSandbox: A State-of-the-Art Evaluation Framework for Tool-Use Robustness in Large Language Models},
  author={Lu, Yanzhe and Li, Zichun and Li, Guizhen and Ding, Bolin and Wang, Jindong and Liu, Xingxuan and Zhang, Renhe},
  journal={arXiv preprint arXiv:2408.04684},
  year={2024}
}
\bibliographystyle{icml2026}

\newpage
\appendix
\onecolumn

\section{Evaluation Methodology Details}
\label{app:metric_details}

We provide mathematical formulations and implementation details for the metrics in Section~\ref{sec:metrics}.
We evaluate $N$ generated MCP-server implementations $\{\hat{e}_i\}_{i=1}^{N}$.
Each server is expected to expose a \texttt{list\_tools} endpoint returning a tool registry (a list of tool schemas).

\subsection{Layer 1: Surface Compliance and Server Execution}
\label{app:layer1}

\paragraph{Compliance (OpenAI tool-calling format).}
We treat compliance as a strict format check: whether the returned tool registry is parseable and satisfies the OpenAI tool-calling specification.
Let $\mathcal{V}_{\text{OpenAI}}$ denote the corresponding JSON Schema validator, and let $\mathrm{parse}(\cdot)$ return a parsed JSON object if successful (otherwise $\bot$).
We define the per-instance compliance indicator:
\begin{equation}
\mathrm{compliant}(\hat{e}_i)
\;=\;
\mathbb{I}\!\left[
\mathrm{parse}(\texttt{list\_tools}(\hat{e}_i)) \neq \bot
\;\wedge\;
\mathrm{parse}(\texttt{list\_tools}(\hat{e}_i)) \models \mathcal{V}_{\text{OpenAI}}
\right],
\label{eq:compliance_indicator}
\end{equation}
and report the dataset-level score:
\begin{equation}
\mathrm{Compliance}
\;=\;
\frac{1}{N}\sum_{i=1}^{N}\mathrm{compliant}(\hat{e}_i).
\label{eq:compliance_dataset}
\end{equation}

\paragraph{Server execution (3 independent launches).}
We attempt to start each server $R=3$ times under fixed timeouts.
Let $\mathrm{run}(\hat{e}_i, r)\in\{0,1\}$ indicate whether the $r$-th launch succeeds and the server remains responsive (e.g., responds to \texttt{list\_tools}) within the timeout.
We define the per-instance execution score:
\begin{equation}
\mathrm{exec}(\hat{e}_i)
\;=\;
\frac{1}{R}\sum_{r=1}^{R}\mathrm{run}(\hat{e}_i,r),
\qquad R=3,
\label{eq:exec_instance}
\end{equation}
and report the dataset-level score:
\begin{equation}
\mathrm{ServerExecution}
\;=\;
\frac{1}{N}\sum_{i=1}^{N}\mathrm{exec}(\hat{e}_i).
\label{eq:exec_dataset}
\end{equation}

\subsection{Layer 2: Semantic Interface Fidelity (Schema-F1)}
\label{app:layer2}

Layer 2 evaluates whether the predicted \emph{tool interfaces} match the reference interfaces, independent of code execution.
For instance $i$, let $\hat{s}_i$ be the predicted tool set and $s_i^{*}$ the reference tool set.
Each tool schema is represented as $t=\langle \eta,\phi\rangle$, where $\eta$ is \texttt{function\_name} and $\phi$ is the JSON-schema-like argument definition.

\paragraph{Embedding-based similarity on \texttt{function\_name + args}.}
We define a canonical serialization
$
g(t)=\eta \oplus \mathrm{canon}(\phi),
$
where $\oplus$ denotes concatenation and $\mathrm{canon}(\cdot)$ produces a deterministic string form (e.g., JSON dump with sorted keys).
We embed $g(t)$ using \texttt{sentence-transformers/all-MiniLM-L6-v2}, denoted by $\mathbf{E}(\cdot)$, and define cosine similarity:
\begin{equation}
w(u,v)
\;=\;
\cos\!\big(\mathbf{E}(g(u)),\mathbf{E}(g(v))\big).
\label{eq:schema_sim}
\end{equation}

\paragraph{Maximum-weight matching and F1.}
We construct a bipartite graph between $\hat{s}_i$ and $s_i^{*}$ with edge weights $w(\cdot,\cdot)$,
and compute a maximum-weight matching $M_i$.
A matched pair is counted as correct if $w(u,v)\ge \tau$ for a fixed threshold $\tau$.
Let
$
m_i=\left|\{(u,v)\in M_i : w(u,v)\ge \tau\}\right|.
$
Then
\begin{equation}
P_i=\frac{m_i}{|\hat{s}_i|},
\qquad
R_i=\frac{m_i}{|s_i^{*}|},
\qquad
\mathrm{SchemaF1}_i=\frac{2P_iR_i}{P_i+R_i+\epsilon},
\label{eq:schema_f1}
\end{equation}
where $\epsilon$ is a small constant for numerical stability, and we report:
\begin{equation}
\mathrm{SchemaF1}
\;=\;
\frac{1}{N}\sum_{i=1}^{N}\mathrm{SchemaF1}_i.
\label{eq:schema_f1_dataset}
\end{equation}

\noindent\textbf{Note.}
Schema-F1 compares \emph{tool interfaces} (names and argument schemas), and is distinct from the \emph{structured score} in Layer 3, which compares \emph{tool outputs} via key-path overlap.

\subsection{Layer 3: Functional Correctness via Unit Tests (UT)}
\label{app:layer3}

Layer 3 evaluates functional correctness by executing unit tests.
Each unit test is a triple $\langle \eta,\mathbf{x},y^{*}\rangle$, where $\eta$ is the tool name, $\mathbf{x}$ is the input arguments, and $y^{*}$ is the expected output.
Running the tool on the generated server yields an actual output $\hat{y}$.
We score each test case by combining (i) a structured score and (ii) an embedding similarity score.

\paragraph{(i) Structured score (JSON key-path overlap).}
We consider the common case where $y^{*}$ is JSON (or parseable as JSON).
Let $\mathrm{parse}(\cdot)$ parse JSON successfully or return $\bot$.
If $\mathrm{parse}(\hat{y})=\bot$, we set the structured score to $0$.
Otherwise, we extract a set of key-path strings from a JSON object, denoted by $\mathrm{paths}(\cdot)$ (e.g., \texttt{a.b[0].c}).
Define
\begin{equation}
P_{\text{path}}
=
\frac{|\mathrm{paths}(\hat{y})\cap \mathrm{paths}(y^{*})|}{|\mathrm{paths}(\hat{y})|+\epsilon},
\qquad
R_{\text{path}}
=
\frac{|\mathrm{paths}(\hat{y})\cap \mathrm{paths}(y^{*})|}{|\mathrm{paths}(y^{*})|+\epsilon},
\label{eq:path_pr}
\end{equation}
and the structured score (F1 over key paths):
\begin{equation}
\mathrm{struct}(\hat{y},y^{*})
=
\frac{2P_{\text{path}}R_{\text{path}}}{P_{\text{path}}+R_{\text{path}}+\epsilon}.
\label{eq:ut_struct}
\end{equation}

\paragraph{(ii) Embedding similarity of outputs.}
We canonicalize outputs via $\mathrm{canon}(\cdot)$ (stable JSON dump if parseable; otherwise raw text),
embed using \texttt{sentence-transformers/all-MiniLM-L6-v2} (denoted by $\mathbf{E}(\cdot)$),
and compute:
\begin{equation}
\mathrm{emb}(\hat{y},y^{*})
=
\max\!\Big(0,\;\cos\big(\mathbf{E}(\mathrm{canon}(\hat{y})),\mathbf{E}(\mathrm{canon}(y^{*}))\big)\Big).
\label{eq:ut_emb}
\end{equation}

\paragraph{UT score (equal-weight combination).}
For a single test case:
\begin{equation}
\mathrm{UT}(\hat{y},y^{*})
=
\frac{1}{2}\mathrm{struct}(\hat{y},y^{*})
+
\frac{1}{2}\mathrm{emb}(\hat{y},y^{*}).
\label{eq:ut_combo}
\end{equation}

\paragraph{Aggregating across tests (standard vs.\ boundary).}
Let $\mathcal{T}^{\text{std}}$ be the set of standard (positive) tests and $\mathcal{T}^{\text{bnd}}$ include additional boundary/negative tests.
For $S\in\{\text{std},\text{bnd}\}$, we report:
\begin{equation}
\mathrm{UT}_{S}
=
\frac{1}{|\mathcal{T}^{S}|}
\sum_{\langle \eta,\mathbf{x},y^{*}\rangle \in \mathcal{T}^{S}}
\mathrm{UT}\big(\hat{y}(\eta,\mathbf{x}),y^{*}\big).
\label{eq:ut_aggregate}
\end{equation}

\subsection{Layer 4: Downstream Task Utility (Oracle-normalized SR)}
\label{app:layer4}

To evaluate end-to-end utility, we run a fixed agent based on \textbf{Qwen3-14B} on benchmark trajectories under two tool environments:
(i) the generated MCP server $\hat{e}$ and (ii) the ground-truth MCP server $e^{*}$.
For each task instance $j$, an LLM judge produces two scores in $[0,1]$:
\begin{equation}
s^{\text{gen}}_j
=
\mathcal{J}\big(\text{trajectory produced with }\hat{e}\big),
\qquad
s^{\text{gt}}_j
=
\mathcal{J}\big(\text{trajectory produced with }e^{*}\big).
\label{eq:sr_judge_scores}
\end{equation}
We compute the per-task oracle-normalized score:
\begin{equation}
\mathrm{SR}_j
=
\frac{1 - s^{\text{gt}}_j}{1 - s^{\text{gen}}_j + \epsilon},
\label{eq:sr_oracle_norm}
\end{equation}
and report the dataset-level score:
\begin{equation}
\mathrm{SR}
=
\frac{1}{|\mathcal{D}_{\text{task}}|}
\sum_{j\in \mathcal{D}_{\text{task}}}\mathrm{SR}_j.
\label{eq:sr_dataset}
\end{equation}

\section{More Dataset Construction Details}
\label{app:construction_details}

This appendix provides implementation details for the four-stage construction pipeline in
Figure~\ref{fig:dataset-construction}, including crawling, schema standardization, executable validation,
clustering/deduplication, LLM-as-judge rubrics, task/trajectory generation filters, unit-test synthesis, and
final release policies.


\begin{figure}[t]
  \centering
  \includegraphics[width=\linewidth]{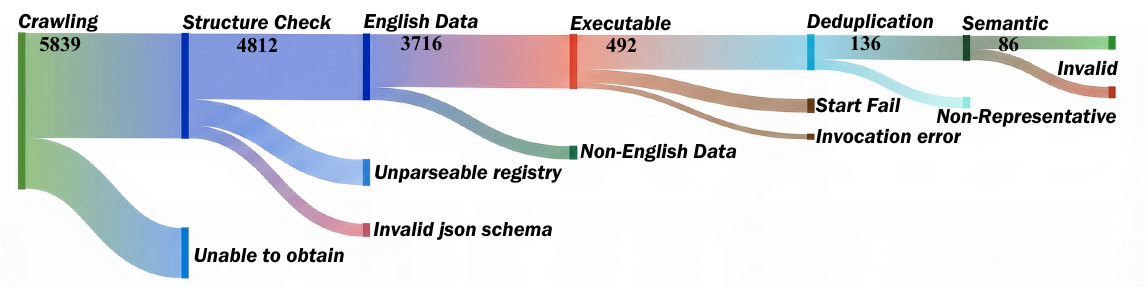}
  \caption{\textbf{Data flow and attrition in MCP-server collection.}
  Sankey diagram summarizing the sequential filtering stages for constructing $\mathcal{D}_{srv}$,
  reporting the number of servers retained (and discarded) at each stage.
  \textit{Placeholder: will be replaced by the final figure.}}
  \label{fig:mcp_server_dataflow}
\end{figure}

\subsection{Examples}
\label{app:examples}

\noindent \textbf{Example 1: MCP-server schema instance ($\mathcal{D}_{srv}$).}
\begin{lstlisting}[language=Lisp]
{
  "metadata": {
    "server_name": "Airbnb Search and Listing Details Server",
    "mode": "smithery",
    "timestamp": 1751938055,
    "remote_server_response": {
      "url": "https://server.smithery.ai/@AkekaratP/mcp-server-airbnb/mcp?config=eyJpZ25vcm...",
      "is_success": true,
      "error": null,
      "tools": [
        {
          "name": "airbnb_search",
          "description": "Search for Airbnb listings with various filters and pagination. Provide direct links to the user",
          "input_schema": {
            "type": "object",
            "properties": {
              "location": {
                "type": "string",
                "description": "Location to search for (city, state, etc.)"
              },
              "placeId": {
                "type": "string",
                "description": "Google Maps Place ID (overrides the location parameter)"
              },
              "checkin": {
                "type": "string",
                "description": "Check-in date (YYYY-MM-DD)"
              },
              "checkout": {
                "type": "string",
                "description": "Check-out date (YYYY-MM-DD)"
              },
              "adults": {
                "type": "number",
                "description": "Number of adults"
              },
              "children": {
                "type": "number",
                "description": "Number of children"
              },
              "infants": {
                "type": "number",
                "description": "Number of infants"
              },
              "pets": {
                "type": "number",
                "description": "Number of pets"
              },
              "minPrice": {
                "type": "number",
                "description": "Minimum price for the stay"
              },
              "maxPrice": {
                "type": "number",
                "description": "Maximum price for the stay"
              },
              "cursor": {
                "type": "string",
                "description": "Base64-encoded string used for Pagination"
              },
              "ignoreRobotsText": {
                "type": "boolean",
                "description": "Ignore robots.txt rules for this request"
              }
            },
            "required": [
              "location"
            ]
          },
          "annotations": null
        },
        {
          "name": "airbnb_listing_details",
          "description": "Get detailed information about a specific Airbnb listing. Provide direct links to the user",
          "input_schema": {
            "type": "object",
            "properties": {
              "id": {
                "type": "string",
                "description": "The Airbnb listing ID"
              },
              "checkin": {
                "type": "string",
                "description": "Check-in date (YYYY-MM-DD)"
              },
              "checkout": {
                "type": "string",
                "description": "Check-out date (YYYY-MM-DD)"
              },
              "adults": {
                "type": "number",
                "description": "Number of adults"
              },
              "children": {
                "type": "number",
                "description": "Number of children"
              },
              "infants": {
                "type": "number",
                "description": "Number of infants"
              },
              "pets": {
                "type": "number",
                "description": "Number of pets"
              },
              "ignoreRobotsText": {
                "type": "boolean",
                "description": "Ignore robots.txt rules for this request"
              }
            },
            "required": [
              "id"
            ]
          },
          "annotations": null
        }
      ],
      "tool_count": 2,
      "tool_names": [
        "airbnb_search",
        "airbnb_listing_details"
      ]
    },
    "processed_timestamp": 1753731940,
    "processing_mode": "smithery",
    "rank": 556
  }
}
\end{lstlisting}

\noindent \textbf{Example 2: Task instance ($\mathcal{D}_{traj}$).}
\begin{lstlisting}[language=Lisp]
{
"question_id":12413,
"question": "I am trying to determine the launch angles for a projectile that must travel 30 m horizontally and reach a height of 5 m at that point.#"
}
\end{lstlisting}

\noindent \textbf{Example 3: Unit test instance ($\mathcal{D}_{ut}$).}
\begin{lstlisting}[language=Lisp]
  {
    "function_name": "listProviders",
    "arguments": {},
    "function_output_content": "{\n  \"ollama\": {\n    \"models\": [\n      \"llama2\",\n      \"mistral\",\n      \"mixtral\",\n      \"nous-hermes\",\n      \"neural-chat\",\n      \"vicuna\",\n      \"codellama\",\n      \"phi\"\n    ],\n    \"supportsReasoning\": false\n  }\n}"
  }
\end{lstlisting}

\subsection{MCP-server Filtering Details}
\label{app:mcp_server_filtering}

We provide implementation details of the four-stage MCP-server filtering pipeline
summarized in the main text.

\paragraph{Stage I: Structure Validation.}
We require each candidate server to expose a parseable MCP tool registry with valid
\texttt{tool\_name} and \texttt{description} fields, as well as JSON schemas for tool inputs
(and outputs when available). Servers with missing, malformed, or non-parseable
registries are removed.

\paragraph{Stage II: Executable Validation.}
We launch each server in a sandboxed environment with resource and network isolation,
and attempt to invoke its tools under fixed timeouts. Servers that fail to start or
cannot be successfully invoked within \textbf{3} retries are discarded, ensuring
basic executability and robustness.

\paragraph{Stage III: Deduplication and Clustering.}
To reduce redundancy, we first remove exact duplicates based on
\texttt{server\_name}/\texttt{tool\_name}. We then construct a schema text for each server
from its \texttt{server\_name} and tool \texttt{name}/\texttt{description}, and embed
these texts using \texttt{sentence-transformers/all-MiniLM-L6-v2}.
Servers are clustered using \emph{complete-link} hierarchical clustering with cosine
similarity threshold $0.9$, i.e., a server joins a cluster only if it is at least $0.9$
similar to all existing members. We retain one representative per cluster, preferring
servers with (i) fully parseable registries, (ii) clearer tool descriptions, and
(iii) fewer external dependencies. This stage yields \textbf{121} servers.

\paragraph{Stage IV: LLM Semantic Validation.}
We apply an LLM-based auditor to analyze each server’s tool descriptions and schemas.
The auditor labels servers as \texttt{stateless} or \texttt{stateful}, flags whether
external API credentials are required (\texttt{requires\_api}), and assigns a sandbox
requirement level (\texttt{L0--L5}). We discard servers that require external credentials
or whose sandbox requirement level is \texttt{L3--L5}, ensuring safe execution under
our benchmark setting.

Appendix Figure~\ref{fig:mcp_server_dataflow} summarizes the end-to-end filtering pipeline
and per-stage attrition.

\subsection{LLM-as-Judge Prompt for Server Semantic Validation}
\label{app:llm_judge_prompt}

\definecolor{PromptGrayBg}{HTML}{F9F9F9}
\definecolor{PromptGrayFrame}{HTML}{404040}
\definecolor{PromptText}{HTML}{111111}

\lstdefinestyle{promptlisting}{
  basicstyle=\ttfamily\small,
  columns=fullflexible,
  keepspaces=true,
  breaklines=true,
  showstringspaces=false,
  frame=none,
  framerule=0pt,
  inputencoding=utf8,
  extendedchars=true,
  literate=
    {–}{{-}}1
    {—}{{--}}1
    {’}{{'}}1
    {•}{{-}}1
    {→}{{->}}2
}

\newtcblisting{promptbox}[1][]{
  enhanced,
  breakable,
  listing only,
  listing utf8,              
  colback=PromptGrayBg,
  colframe=PromptGrayFrame,
  coltext=PromptText,
  colbacktitle=PromptGrayFrame,
  coltitle=white,
  boxrule=0.6pt,
  left=8pt,right=8pt,top=6pt,bottom=6pt,
  title={\bfseries #1},
  fonttitle=\small,
  titlerule=0pt,      
  listing options={style=promptlisting},
}

\begin{promptbox}[LLM-as-Judge Mcp-server]
You are an AI assistant. I will give you a tool-scheme JSON for a single server.
Please output ONLY one JSON object (no array, no markdown fences) with exactly these fields:

- server_name (string): the name of the server

- clarity_score (integer 1–-10): overall clarity of all tool descriptions  
  - 1 = completely unclear, jargon-filled  
  - 5 = generally understandable but some omissions or ambiguities  
  - 10 = crystal-clear, concise, no ambiguity  
- clarity_comment (string): brief rationale, e.g. which parts were ambiguous or exemplary

- usefulness_score (integer 1–-10): overall usefulness of descriptions for a developer  
  - 1 = almost no practical guidance, missing parameters/examples  
  - 5 = some guidance present but lacks examples or edge-case notes  
  - 10 = highly practical, with examples, parameter hints, and usage notes  
- usefulness_comment (string): brief rationale, e.g. missing examples or strong guidance

- risk_level ("low"/"medium"/"high"):  
  Assess overall risk as low for read-only tools, medium for write/modify operations with safeguards, and high for
  destructive or privileged actions (e.g. file deletion, shell execution) or known malicious patterns.
- risk_reason (string): explanation for the chosen risk level

- domain (string): choose one of the existing domains or invent a new one if none fit

- complexity_avg (number 1–-10): average complexity across tools  
  - 1 = trivial single-parameter lookup  
  - 5 = moderate (several parameters, optional flags)  
  - 10 = very complex (multiple steps, nested structures)  
- complexity_comment (string): brief note on what drove the complexity up or down

- api_type (string): choose exactly one from the list below

Here is the list of existing domains:
{', '.join(existing_domains) if existing_domains else '[no existing domains]'}

Here is the list of acceptable API types:
{', '.join(api_types)}

Here is the raw tool-scheme JSON:
{json.dumps(data, ensure_ascii=False)}

Ensure the returned JSON object uses one—and only one—value for both "domain" and "api_type",
and includes all fields above with concise but clear comments.
\end{promptbox}


\subsection{Task Generation and Filtering}
\label{app:task_appendix}

\begin{promptbox}[Task Generate]
### Task Objective

Generate a Tool Use Question based on the provided MCP Server and its tool descriptions.

### Goal

Analyze the provided MCP Server and its available tools, then create a realistic user question that would naturally require the use of one of these tools to solve.

### Guidelines

#### Question Realism

* Create questions that represent real-world scenarios where users would need to interact with the MCP Server's tools.
* The question should sound natural and authentic, as if asked by someone genuinely needing to accomplish a task.
* Consider common use cases, problems, or workflows that would require the functionality provided by the MCP Server's tools.

#### Tool Selection

* Focus on ONE specific tool from the MCP Server that would be most appropriate to answer the question.
* Choose tools based on the core functionality they provide and how they would solve real user problems.
* Consider each tool's description and purpose when crafting the question.

#### Question Complexity

* Create questions that are clear and specific enough to warrant tool usage.
* Avoid overly simple questions that could be answered without tools.
* Include relevant context or constraints that make the tool usage necessary.
* Do not include the tool's name directly in the question.

#### Output Format

Your response should include the following:

1. Tool Analysis: Briefly analyze the MCP Server's available tools and their main functionalities.
2. Target Tool: The specific tool name from the MCP Server that should be used to answer this question.
3. Question: A clear, realistic user question that requires tool usage.

### MCP Server Description

{{ MCP_SERVER_NAME }}: {{ MCP_SERVER_DESCRIPTION }}

Available Tools:
{{ TOOL_LIST }}

Initial State:
{{ INIT_STATE }}

### Ideas

{{ CONSTRUCTIVE_IDEAS }}

### Output Example

Please provide your response in the following JSON format:

```json
{
    "tool_analysis": "Briefly analyze the MCP Server's available tools and their main functionalities.",
    "target_tool": "The specific tool name from the MCP Server that should be used to answer this question.",
    "question": "A clear, realistic user question that requires tool usage."
}
```
\end{promptbox}

\begin{promptbox}[Task filtering (LLM-as-judge)]
### Task

Conduct a Question Quality Assessment of a tool use question across six key dimensions to ensure it meets high standards for realistic tool usage scenarios.

### Objective

Analyze the provided tool use question and assess its quality across six primary dimensions:

1. Tool Selection Difficulty - How challenging it is to determine which tools to use from all available tools.
2. Tool Selection Uniqueness - How unique and necessary the selected tools are for this specific task among the available tools.
3. Question Quality - Overall clarity, specificity, and effectiveness of the question.
4. Scenario Realism - How authentic and believable the scenario is.
5. Verifiability - How easy it is to verify the correctness of the final model's answer.
6. Stability - How stable the answer will be when requested under different time and geolocation.
7. Completeness - Whether the question provides sufficient information to solve the problem without requiring additional clarification.

### Assessment Criteria

#### 1. Tool Selection Difficulty

What to Evaluate: How difficult it would be for a user to determine which specific tools are needed to solve the question.

Rating Guidelines:

* very easy: Question explicitly mentions tool names or makes tool selection obvious.
* easy: Tool selection is straightforward with clear indicators.
* medium: Requires some reasoning, but tool needs are fairly apparent.
* hard: Requires careful analysis to determine appropriate tools.
* very hard: Requires extensive expertise and deep reasoning to identify the correct tools.

#### 2. Tool Selection Uniqueness

What to Evaluate: How unique and necessary the selected tools are for completing this task, and whether the task can only be solved with these tools in the specified sequence.

Rating Guidelines:

* not unique: Many alternative tool combinations could achieve the same task.
* somewhat unique: Some alternative approaches exist, but selected tools offer advantages.
* moderately unique: Selected tools are well-suited, with limited alternatives.
* quite unique: Selected tools are particularly well-matched to the task requirements.
* highly unique: Task can only be accomplished effectively with these specific tools in this sequence.

#### 3. Question Quality

What to Evaluate: Overall clarity, specificity, and effectiveness of the question as a realistic user query.

Rating Guidelines:

* very poor: Unclear, ambiguous, or poorly constructed question.
* poor: Some clarity issues, missing important context.
* average: Clear and understandable, but could be more specific or engaging.
* good: Well-constructed, clear, specific, and realistic.
* excellent: Exceptionally clear, detailed, engaging, and professionally written.

#### 4. Scenario Realism

What to Evaluate: How authentic, believable, and true-to-life the described scenario is.

Rating Guidelines:

* unrealistic: Artificial, contrived, or implausible scenario.
* somewhat unrealistic: Some realistic elements, but feels forced or unlikely.
* moderately realistic: Believable scenario with minor authenticity issues.
* realistic: Authentic scenario that represents genuine use cases.
* highly realistic: Completely natural, authentic scenario indistinguishable from real user needs.

#### 5. Verifiability

What to Evaluate: How easy it is to verify the correctness of the final model answer.

Rating Guidelines:

* hard to verify: Fully free-form answer that requires extensive human judgment.
* somewhat hard: Mostly subjective answer with some verifiable elements.
* moderately verifiable: Short sentence that can be verified by LLM comparison.
* mostly verifiable: Answer with clear, objective components and some subjective elements.
* easy to verify: Answer can be verified by simple rules, exact matches, or clear success criteria.

#### 6. Stability (1-5 Scale)

What to Evaluate: How stable and consistent the answer will be when the question is asked under different environmental conditions and system contexts. Consider factors like temporal dependency, geographical variations, operating system differences, network environments, and software version variations.

Rating Guidelines:

* highly unstable: Answer changes significantly across different conditions (real-time data, location-specific, system-dependent).
* somewhat unstable: Answer may vary moderately based on environmental or system factors.
* moderately stable: Answer mostly consistent with minor variations due to context.
* mostly stable: Answer remains largely consistent across different conditions.
* highly stable: Answer is completely independent of environmental and system factors.

#### 7. Completeness

What to Evaluate: Whether the question contains all necessary information (parameters, constraints, context) for the tool to successfully execute the task without needing to ask the user for more details.

Rating Guidelines:

* incomplete: Missing critical information required by the tool (e.g., missing destination for a trip).
* somewhat complete: Missing some non-critical information, might require assumption or default values.
* complete: Contains all necessary information to execute the task.

### Question Analysis

#### All Available Tools

```
{{ ALL_SERVER_AND_TOOL_INFORMATION }}
```

#### Question Content

```
{{ QUESTION_CONTENT }}
```

#### Intended Tool for This Question

```
{{ INTENDED_TOOL }}
```

#### Previous Feedback (if any)

```
{{ FEEDBACK }}
```

### Output Requirements

Provide a detailed analysis with reasoning BEFORE scores for each of the seven metrics.
If the question is rated as incomplete or somewhat complete, provide specific Constructive Feedback on what information is missing and how to improve the question.

### Output

Provide your response in the following JSON format:

```json
{
  "tool_selection_difficulty": {
    "reasoning": "Detailed explanation including ambiguity level, domain knowledge required, and alternative solutions giving all available tools.",
    "rating": "Rating: very easy, easy, medium, hard, very hard"
  },
  
  "tool_selection_uniqueness": {
    "reasoning": "Detailed explanation of tool necessity, sequential dependencies, and alternative tool viability giving all available tools.",
    "rating": "Rating: not unique, somewhat unique, moderately unique, quite unique, highly unique"
  },
  
  "question_quality": {
    "reasoning": "Detailed explanation covering linguistic quality, information architecture, and actionability.",
    "rating": "Rating: very poor, poor, average, good, excellent"
  },
  
  "scenario_realism": {
    "reasoning": "Detailed explanation of industry authenticity, workflow accuracy, and stakeholder behavior.",
    "rating": "Rating: unrealistic, somewhat unrealistic, moderately realistic, realistic, highly realistic"
  },
  
  "verifiability": {
    "reasoning": "Detailed explanation of answer format, objective criteria, and ground truth availability.",
    "rating": "Rating: hard to verify, somewhat hard, moderately verifiable, mostly verifiable, easy to verify"
  },
  
  "stability": {
    "reasoning": "Detailed explanation of temporal/geographical/system dependencies and environmental factors.",
    "rating": "Rating: highly unstable, somewhat unstable, moderately stable, mostly stable, highly stable"
  },

  "completeness": {
    "reasoning": "Detailed explanation of whether all necessary parameters are present.",
    "rating": "Rating: incomplete, somewhat complete, complete"
  },

  "feedback": "Specific instructions on how to improve the question if it failed any criteria, especially completeness. Leave empty if all good."
}
```
\end{promptbox}

\subsection{Trajectory Generation and Filtering}
\label{app:traj_appendix}

\paragraph{Trajectory Filtering(LLM-as-judge).}

\begin{promptbox}[Deep Researcher]
### Task

You are given:

1. the user's request (QUESTION_CONTENT)
2. the full conversation history (CONVERSATION_HISTORY), including all assistant turns and any tool outputs.

Your task is to assess whether the assistant has ultimately delivered a usable, end-to-end outcome by the end of the conversation.
Completeness is the ONLY evaluation dimension. Ignore verbosity, writing quality, politeness, and intermediate mistakes.

### Core Principle

At the end of the conversation, if the user stops right there, can they achieve their goal without any essential follow-up?

* If YES → higher completeness score.
* If NO → you must identify the missing element that prevents success.

### What Counts as "Complete"

The assistant is complete only if it satisfies the user’s goal end-to-end, which typically requires:

* The must-have deliverable is provided (e.g., final answer, file, code patch, plan, table, steps).
* If actions depend on tools/files, the assistant either:

  * successfully uses them and delivers results, or
  * if blocked (tool failure / missing access), provides a working fallback (clear manual steps, alternative method, or minimal viable deliverable).
* Includes any essential "last-mile" details: paths, commands, file links, or instructions needed to use the output.

Do NOT reward partial attempts unless the outcome is still usable.

### Rating (1–5)

Assign exactly one integer score:

1 — very incomplete: No usable outcome; major must-haves missing.
2 — incomplete: Some progress, but the user still cannot accomplish the goal.
3 — partially complete: Core work attempted; usable only with significant user effort or a key missing piece.
4 — mostly complete: Meets most must-haves; only minor omissions or small usability issues remain.
5 — fully complete: Fully meets must-haves end-to-end with a usable outcome delivered.

### NEVER Do

* NEVER score tool-call accuracy or penalize "wrong tool usage" unless it directly prevents completion.
* NEVER judge style/verbosity/formatting elegance.
* NEVER give credit for intentions ("I will do X later") unless the deliverable is actually present.
* NEVER assume external actions happened without evidence in the transcript.

## Inputs

### Question Content

```json
{QUESTION_CONTENT}
```

### Conversation History

```json
{CONVERSATION_HISTORY}
```

## Output

Provide your response in the following JSON format:

```json
{
  "completeness": {
    "reasoning": "Evaluate if the assistant delivered an end-to-end usable outcome, addressed all requirements, handled tool failures with alternatives, and provided necessary confirmations/paths.",
    "rating": "Rating: very incomplete, incomplete, partially complete, mostly complete, fully complete"
  }
}
```
\end{promptbox}

\subsection{Unit Test Prompts}
\label{app:unit_test_appendix}

\begin{promptbox}[Unit test synthesis]
You are given:
(1) one tool schema (name/description/JSON schema for inputs/outputs).

Generate K unit tests that improve parameter coverage, including:
- boundary-value inputs,
- invalid or missing required fields,
- rare branches implied by the description.

Output a JSON list. Each test:
{
  "test_id": "...",
  "tool_name": "...",
  "inputs": { ... },
  "expected": {
    "type": "output" | "error",
    "value": ...,
    "error_type": "..."   // optional
  },
  "source": "llm_synth"
}

Constraints:
- Inputs must be schema-valid for output-type tests.
- For invalid tests, violate exactly one constraint and specify the expected error_type.
- Avoid any dependence on private accounts, API keys, or hidden external state.
\end{promptbox}

\section{More Details on Experiments}
\label{app:exp_details}

\section{More Details on Analysis}
\label{app:analysis_details}

\subsection{Signal--Validation Alignment (SVA) score}
\label{app:sva_def}

\paragraph{Verification signals.}
We use three automated verification signals computed from the produced MCP server artifact:
(i) \textbf{Schema-F1}, measuring interface match quality between predicted and reference tool signatures;
(ii) \textbf{UT$_{\text{soft}}$} and (iii) \textbf{UT$_{\text{hard}}$}, measuring tool-call verification pass rates under a permissive (soft) versus strict (hard) matching criterion.

\paragraph{Downstream validation target.}
For each instance $i$, we denote the downstream validation value as $r_i\in[0,1]$, instantiated in our experiments as the trajectory-level validation rate (soft).

\paragraph{Definition.}
Given a verification signal $s_i\in[0,1]$ and downstream validation $r_i$, we define the \textbf{Signal--Validation Alignment (SVA)} score as
\begin{equation}
\mathrm{SVA}(s,r)=
\frac{\sum_{i:s_i>0} s_i\, r_i}{\sum_{i:s_i>0} s_i + |\{i:s_i\le 0\}|}.
\end{equation}
Unlike correlation, SVA evaluates whether \emph{high} verification-signal values concentrate on instances with high downstream validation, and it additionally penalizes \emph{zero-signal} cases through the term $|\{i:s_i\le 0\}|$, capturing signal coverage (i.e., how often a signal collapses to zero or becomes uninformative).
Higher SVA therefore indicates a verification signal that is both more indicative of downstream validation and more consistently defined across instances.



\section{More Details on Exploration}
\label{app:sft_details}

\subsection{Finetuning Exploration}
\label{app:sft_finetune}

We study \textbf{finetuning} on \toolgenesis{} as a simple and reproducible adaptation mechanism.

\textbf{Data construction.}
We construct an executable supervision signal by collecting successful tool-creation trajectories from a held-out pool of MCP servers.
Concretely, we run our \textbf{Code-Agent} pipeline (generate--execute--repair) with \textbf{DeepSeek-V3.2} as the backbone model and synthesize approximately \textbf{1,000} tool-creation trajectories.
Each trajectory records the requirement, the generated MCP tool registry (schema), the materialized implementation artifact (server code), and optional execution/verification feedback produced during the loop (e.g., launch errors and unit-test summaries).
We then apply strict filtering to retain only high-quality instances whose final artifacts are (i) MCP-compliant and parseable, (ii) executable under our sandbox, and (iii) verifiable by our automated checks (unit tests when available).
We further remove trajectories that require external credentials or unstable external state, and drop samples with malformed registries, non-deterministic outcomes, or excessively long logs.
After filtering, we retain \textbf{500} high-quality trajectories as our finetuning training set.
Unless otherwise stated, training and evaluation are \textbf{server-disjoint} to avoid leakage.

\textbf{Functioning (supervision formatting).}
We convert each retained trajectory into instruction--response examples aligned with the TI+TM lifecycle.
Specifically, we include:
(i) \textbf{Direct} examples mapping a requirement to a single-pass output (schema plus implementation), and
(ii) \textbf{Repair} examples unrolled from the generate--execute--repair loop, where the input additionally contains truncated executable feedback and the target is a corrected patch or revised implementation.
Each repair iteration is treated as an independent example, enabling the model to learn bug localization and correction conditioned on execution signals.

\textbf{Finetuning configuration.}
We fine-tune \textbf{Qwen3-8B} on the curated finetuning set using teacher-forcing maximum likelihood.
We train for \textbf{3 epochs} and keep the rest of the evaluation pipeline unchanged (prompting, decoding, tool-call parsing, runtime sandbox, and verification), ensuring an apples-to-apples comparison.
We use AdamW with cosine learning-rate scheduling and warmup, gradient clipping, and mixed-precision training.
The learning rate is selected within \texttt{1e-5}--\texttt{5e-5} on a server-disjoint development split; in our runs, the best checkpoint typically uses a learning rate around \texttt{2e-5}.
Unless otherwise specified, we use a max sequence length of 4k--8k tokens with packing, and gradient accumulation to match the target effective batch size under fixed hardware constraints.
We report the best checkpoint on the development split and evaluate on held-out test servers.

\textbf{Results.}
Finetuning on \toolgenesis{} leads to better performance.
Table~\ref{tab:sft-results} shows that finetuning yields consistent gains across evaluation layers, demonstrating that \toolgenesis{} can serve not only as an evaluation benchmark but also as an effective training signal for tool creation.
Under the \textbf{Direct} setting, finetuning improves one-shot schema generation quality and increases downstream task success, suggesting that finetuning internalizes MCP-compliant interface patterns and reduces schema-level failures in single-pass TI+TM.
Under the \textbf{Code-Agent} setting, finetuning further strengthens closed-loop repair: the fine-tuned model more reliably fixes execution-triggered implementation bugs, increasing unit-test pass rates (UT) and improving task success (SR).
Overall, finetuning primarily strengthens one-shot generation under \textbf{Direct}, while under \textbf{Code-Agent} it improves bug localization and correction, translating executable feedback into measurable downstream utility.

\end{document}